\documentclass[12pt]{article}
\usepackage{amssymb,latexsym}
\usepackage{braket}
\pdfoutput=1

\usepackage{fullpage}

\usepackage[english]{babel}

\usepackage[compress]{cite}

\usepackage[intlimits]{amsmath}

\usepackage{graphicx}
\usepackage{epsfig}
\usepackage{color}

\usepackage{ifpdf}

\ifpdf

\usepackage[pdftex]{hyperref}
\else
\usepackage[hypertex,ps2pdf,dvips,colorlinks,urlcolor=blue,citecolor=blue]{hyperref}
\fi

\def\be{\begin{equation}}
\def\ee{\end{equation}}

\def\bea{\begin{eqnarray}}
\def\eea{\end{eqnarray}}

\begin{document}

\thispagestyle{empty}

 \begin{center}
{\LARGE\bf Quantum quenches from excited states\\
 in the Ising chain}
\\[2.3em]
\bigskip

{\large \bf Leda Bucciantini$^{1}$, M\'arton Kormos$^{1,2}$, Pasquale Calabrese$^{1}$}
\null

\noindent 

{\it 
$^1$ Dipartimento di Fisica dell'Universit\`a di Pisa and INFN, 56127 Pisa, Italy\\
$^2$ MTA-BME ``Momentum'' Statistical Field Theory Research Group,\\
1111 Budapest, Budafoki \'ut 8, Hungary
}

\date{\today}

\end{center}

\begin{abstract}

We consider the non-equilibrium dynamics after a sudden quench of the magnetic field in the transverse field Ising chain
starting from excited states of the pre-quench Hamiltonian. 
We prove that stationary values of local correlation functions can be described by the generalised Gibbs ensemble (GGE). 
Then we study the full time evolution of the transverse magnetisation by means of stationary phase methods. 
The  equal time two-point longitudinal correlation function is analytically derived for a particular class of excited states
for quenches within the ferromagnetic phase, and studied numerically in general.  
The full time dependence of the entanglement entropy of a block of spins is also obtained analytically for the 
same class of states and for arbitrary quenches. 

\end{abstract}


\newpage

\section{Introduction}

Recent experiments in the field of ultra-cold atoms \cite{kww-06,tetal-11,cetal-12,getal-11,shr-12} triggered
an intense theoretical activity aimed to understand the unitary non-equilibrium evolution of 
isolated many-body quantum systems.
A situation which attracted a lot of interest is that of a sudden quench of a Hamiltonian 
parameter such as an external magnetic field \cite{revq}. 
The most remarkable results that emerged from these theoretical and experimental investigations are probably the following two:
(i) there is a light-cone like spreading of correlations following a quantum quench \cite{cetal-12,cc-06,cc-07,lr-72,var-lc} and
(ii) expectation values of {\it local} observables generically approach stationary values at late times in the thermodynamic limit,
although the whole system is always in a pure state. 

Rather amazingly, these stationary values can be predicted by a statistical ensemble without solving the complicated 
non-equilibrium dynamics. 
For non-integrable models the appropriate statistical ensemble is expected to be the standard Gibbs one 
with an effective temperature fixed by the value of the energy in the initial state \cite{nonint}. 
For integrable models, the existence of non-trivial local conservation laws strongly constrains the dynamics 
and the stationary values are expected to be described by a generalised Gibbs ensemble (GGE) \cite{GGE}.
The most convincing evidence supporting this scenario comes from the exact solution of models 
that can be mapped to free fermions or bosons for which the full dynamics can be obtained analytically. 
Among these models a crucial role has been played by the transverse field Ising chain 
which is probably the most intensively investigated model in the quench literature 
\cite{bm-70,ir-00,sps-04,cc-05,fc-08,s-08,rsm-10,ir-10,cef,gcg-11,se-12,cef-i,cef-ii,eef-12,fe-13a,f-13,g-13,f-14}, 
although many other free-like models have been studied as well.
\cite{c-06,bs-08,cdeo-08,scc-09,mc-12,csc-13,kcc-13,rs-13,ckc-13}.
For integrable interacting models, i.e. with a non-trivial scattering matrix between quasi-particle excitations,
the exact solution of the quench dynamics is still an outstanding problem \cite{ce-13}, 
but recently it has been possible in some instances to construct the GGE and compute explicitly a few observables 
\cite{ce-13,fm-10,p-11,mc-12b,ck-12,sfm-12,fe-13b,p-13b,kscci-13,m-13,p-13a,fp-13,nwbc-13,a-12,stm-13,bdwc-14,fcec-13}
and also to check these predictions numerically \cite{fcec-13}. 
Clearly, for non-integrable models the evidence comes only from 
numerical \cite{kla-06,rdo-08,r-09,roux-09,bkl-10,bch-11,ccr-11,bdkm-12,crfss-12,rs-12,sks-13,wrdk-14,r-14,ekmr-14}
or experimental \cite{kww-06,tetal-11} investigations.

However, most of (if not all) these studies focused on the evolution starting from the 
ground-state of a given pre-quench Hamiltonian. 
While these initial states should include all the most relevant experimental situations, 
it is natural to wonder how general the conclusions drawn on their basis are. 
Indeed, ground states of local Hamiltonians are not at all generic because their entanglement entropy 
(i.e. the von Neumann entropy of the reduced density matrix of a subsystem) 
satisfies an area law \cite{area}  (i.e. it scales with the area of the surface of the subsystem) 
or at most it has multiplicative logarithmic corrections, while generic states satisfy a volume law.
This important difference is at the heart of the machinery of the so-called tensor network 
techniques used to effectively simulate many-body quantum systems (see e.g. Refs. \cite{cv-09} as reviews).
Somewhat more general conclusions can be drawn by considering a quantum quench in which 
the initial state is not a ground-state, but an {\it excited state of the pre-quench Hamiltonian},
which generically follows a volume law for the entanglement entropy \cite{afc-09} 
rather than an area law. 

In order to give a first answer to this question, in this paper we consider the quench dynamics in  
the prototype of the models mappable to free fermions, 
namely the transverse field Ising chain, which, in spite of its simplicity and the fact that it is exactly solvable, 
represents a crucial paradigm for quantum critical behaviour. 
Furthermore, while the model admits a representation in terms of free fermions, the spin variables are non-local 
with respect to the fermionic degrees of freedom, a property which renders it less trivial than free particle 
systems and, at the same time, an ideal testing ground for studying the relaxation 
for which locality is an essential feature.

The manuscript is organised as follows. 
In Sec. \ref{diago} we report all the preliminary information to set up the calculations, i.e.
we introduce the model, its diagonalisation, the quench protocol and the observables we will study. 
In Sec. \ref{equivalence} we show that in the long time limit (taken after the thermodynamic one) 
the system can be described by a GGE.  
In Sec. \ref{magne} we study the time evolution of the transverse magnetisation, while in 
the following Sec. \ref{longe} we consider the two-point longitudinal correlation function.
In Sec. \ref{enta} we turn to the time evolution of the entanglement entropy.
In the last section we draw our conclusions and in two appendices we report some additional technical details.

\section{The model, the quench protocol, and the observables}
\label{diago}

We consider here the non-equilibrium dynamics of the transverse field Ising chain with Hamiltonian 
\begin{equation}\label{h_Ising chain}
H_I=-\frac{1}{2} \sum_{j=1}^{N}{[\sigma^x_j\sigma^x_{j+1}+h\sigma_j^z]},
\end{equation} 
where $\sigma^\mu_m$, $\mu=x, y,z$ is the Pauli matrix at site $m$ of the chain, $h$ is the transverse field, 
and we impose periodic boundary conditions $ \sigma^\mu_{N+1}=\sigma^\mu_1$.
The transverse field Ising chain is a crucial paradigm of quantum phase transitions \cite{sach-book} because  at zero temperature 
and in the thermodynamic limit $N\to\infty$ it exhibits ferromagnetic ($h < 1$) and paramagnetic ($h > 1$) phases, 
separated by a quantum critical point at $h_c = 1$.

\subsection{The diagonalisation of the Hamiltonian}

The diagonalisation of Hamiltonian (\ref{h_Ising chain}) is presented in several textbooks (see e.g. \cite{sach-book}), but 
we repeat it here to make the paper self consistent.
The first step in diagonalising the Hamiltonian (\ref{h_Ising chain}) is to introduce a set of spinless fermion 
annihilation and creation operators through the {\it non-local} Jordan-Wigner transformation
\begin{equation}\label{J_W_t}
c_l=\left(\prod_{m<l}\sigma_m^z\right)\frac{\sigma_l^x-i\sigma_l^y}{2},\quad\quad  c_l^\dagger=\left(\prod_{m<l}\sigma_m^z\right)\frac{\sigma_l^x+i\sigma_l^y}{2},
\end{equation}
where the operators $c_l$, $c_l^\dagger$ satisfy anti-commutation relation 
\begin{equation}
\{c_l^\dagger,c_m\}=\delta_{lm}, \quad\quad \left\{c_l,c_m\right\}=0.
\end{equation}
Then the Fourier modes $d_k$ are defined as
\begin{equation}\label{def_d}
d_k=\frac{1}{\sqrt{N}}\sum_{j=1}^{N}{c_l e^{-i(2\pi kj)/N}}.
\end{equation}
Since the transformation is unitary, also the $d_k$ operators satisfy anti-commutation relations.
In terms of $d_k$ operators, the Hamiltonian is
\begin{equation}\label{H_ds}
H_I=\sum_{k}\left(\left[h-\cos\frac{2\pi k}{N}\right]d^\dagger_k d_k-\frac{i}{2}\sin{\frac{2\pi k}{N}}\left[d_{-k} d_k+d^\dagger_{-k} d^\dagger_k\right]-\frac{h}{2}\right),
\end{equation}
where the sum over the modes $k$ runs over integers or half-integers depending on the parity of the fermion 
number, see e.g. the Appendix of Ref. \cite{cef-i} for a detailed discussion.
This is however superfluous for our goals since we will only consider the thermodynamic limit 
in which the momentum becomes continuous. 
 
It is now necessary to apply one more unitary transformation to cast the Hamiltonian in a diagonal form: 
the transformation is a  Bogoliubov rotation which takes the fermionic operators $d_k$, $d^\dagger_k$ 
to the new ones $b_k$, $ b^\dagger_k$
\begin{equation}\label{bogo_diag}
b^\dagger_k=u_k  d^\dagger_k+i v_k  d_{-k}, \quad\quad  b_k=u_k  d_k-i v_k  d^\dagger_{-k},
\end{equation}
where the coefficients $u_k$ and $v_k$ are chosen in such a way as to make $H_I$ diagonal
\begin{equation}\label{u}
u_k=\cos{\left(\theta_k/2\right)}, \quad\quad v_k=\sin{\left(\theta_k/2\right)},
\end{equation}
and the angle $\theta_k$ is defined by the relation
\begin{equation}\label{theta}
\tan{\theta_k}=\frac{\sin{(2\pi k/N)}}{\cos{(2\pi k/N)}-h}.
\end{equation}
Again, the unitarity of the transformation ensures the validity of the usual anti-commutation relation for the $b$ operators. 
Moreover, the characteristic of (\ref{bogo_diag}) of mixing only the modes $k$ and $-k$ of the operators $d$ 
allows us to re-write the Bogoliubov transformation in a 
more compact way as a rotation in a $2 \times 2$ Hilbert space
\begin{equation}
 D_k=R_x(\theta_k) B_k,
 \label{bog2}
\end{equation}
where $D_k$ and $ B_k$ are the two-component vectors
\begin{equation}
B_k=\left(
\begin{array}{c}
 b_k\\
 b^\dagger_{-k}
\end{array}
\right), \quad\quad
 D_k=\left(
\begin{array}{c}
 d_k\\
 d^\dagger_{-k}
\end{array}
\right),
\end{equation}
and the matrix $R_x(\alpha)$ is a special case of
\begin{equation}
R_\mu(\alpha)=\cos{\frac{\alpha}{2}}+i\sigma^\mu\sin{\frac{\alpha}{2}}.
\end{equation}

The Hamiltonian can be written in terms of the Bogoliubov quasi-particles as
\begin{equation}\label{H_diag_preq}
H_I=\sum_k{\epsilon_k\left(b^\dagger_k  b_k-\frac12\right)},
\end{equation}
where the one-particle dispersion relation is
\begin{equation}
\epsilon_k=\sqrt{\left(h-\cos\frac{2\pi k}{N}\right)^2+\sin^2\left(\frac{2\pi k}{N}\right)}\,.
\end{equation}
In the thermodynamic limit, the momentum $\varphi_k=2\pi k/N$ becomes a continuos variable $\varphi$ 
living in the interval $\varphi\in[-\pi,\pi]$.
In the following we will often use $k$ instead of $\varphi$, being sure that it will always be clear whether 
we refer to continuous or discrete momenta. 

\subsubsection{Other useful fermionic operators}

For the calculation of correlation functions it is convenient to introduce some other sets of fermionic 
operators. 
First, we introduce the Majorana fermions
\begin{equation}
A^x_j=c_j^\dagger+ c_j, \quad\quad A_j^y=i(c_j-c^\dagger_j) 
\end{equation}
which satisfy the algebra 
\begin{equation}
\{A_l^x, A_n^x\}=2\delta_{ln}, \quad   \{A_l^y, A_n^y\}=2\delta_{ln},\quad \{A_l^x, A_n^y\}=0.
\end{equation}
While the spin operator $\sigma_n^z$ is local in terms of these Majorana fermions, $\sigma_n^z=i A_n^y A_n^x$, the operator $\sigma_n^x$ has the non-local representation
\begin{equation}
\label{eq:sigmax}
\sigma_n^x=\prod_{j=1}^{n-1}(i A^y_j A^x_j) A^x_n,
\end{equation}
which, as we shall see, is particularly useful in the calculation of real space correlation functions and the entanglement entropy. 

The operators $A^x_j$ and $A^y_j$ can be collected together in two different ways. 
On the one hand, one can introduce a single set of operators at the price of doubling their number per site
in the following way
\be
A_{2j-1}=A^y_j\,,\qquad A_{2j}=A^x_j\,,
\ee
and they satisfy the algebra 
\begin{equation}
\{A_a, A_b\}=2\delta_{ab}.
\end{equation} 
On the other hand, $A^x_j$ and $A^y_j$ can be collected in a  two-component vector operator 
\begin{equation}\label{Omegar}
\Omega_j=\left(
\begin{array}{c}
i A^y_j  \\
A_j^x 
\end{array}
\right).
\end{equation}
We will denote the Fourier transform of $A^x_j$ with $\omega_k^+$
and the Fourier transform of $iA^y_j$ with $\omega_k^-$, and, 
with a slight abuse of notation, we will denote the Fourier transform  of the vector (\ref{Omegar})
simply as  $\Omega_k$:
\begin{equation}\label{Omega}
\Omega_k=\left(
\begin{array}{c}
\omega_k^-  \\
\omega_k^+ 
\end{array}
\right) =\frac{1}{\sqrt{N}}\sum_{j=1}^{N}{ e^{-i(2\pi kj)/N}}
\left(
\begin{array}{c}
i A^y_j  \\
A_j^x 
\end{array}
\right).
\end{equation}

\subsection{The exact spectrum}

The ground state of $H_I$ is the vacuum of Bogoliubov operators, i.e. it is annihilated by all $b_k$.
Its energy is $E_{GS}=-\frac12 \sum_k \epsilon_k$. 
However, the exact diagonalisation of the model gives not only the ground state 
properties but all the eigenstates and their energies.
In the basis of free fermions, the excited states are classified
according to the occupation numbers of the single-particle basis. 
An eigenstate can be then written as 
\be
|m_k\rangle \equiv\prod_{k} (b^\dag_k)^{m_k}|0\rangle\,,
\qquad {\rm with\; energy}\;\; E_{m_k}-E_{GS}=\sum_{k} m_k \epsilon_k\,,
\label{ES}
\ee 
where $m_k=0,1$ is a characteristic function of the state representing the set of occupied momenta,
i.e. $m_k=1$ if the momentum $k$ is occupied and $m_k=0$ if not.  
All observables can be written in terms of the characteristic function as for example the energy 
in Eq. (\ref{ES}).
While in finite system the characteristic function can assume only the values $0$ and $1$, 
in the thermodynamic limit it becomes an arbitrary function $m(\varphi)$ of $\varphi\in [-\pi,\pi]$ with the 
restriction to be in the interval $[0,1]$. This function $m(\varphi)$ is a coarse-grained version of $m_k$, 
see e.g. \cite{afc-09} for specific examples.  
The ground state corresponds to $m_k=0$ for every $k$.

\subsection{The quench protocol}

The time dependence of the system after a quench of the transverse field starting from the 
ground state has been studied extensively in a series of works, as reported in the introduction.
Here we are interested in the case when the initial state is an excited state of the pre-quench Hamiltonian, i.e. 
for $t<0$ the system is in an excited state of the Hamiltonian $H_I$ with field equal to $h_0.$
At time $t=0$ the value of the field is suddenly quenched to $h\neq h_0$ and all the following 
time evolution is governed by this new Hamiltonian.
We will denote by $b'_k$ and $ b_k$ the fermionic mode operators that diagonalise the Hamiltonian 
with $h_0$ and $h$, respectively. 
Similarly, primed symbols will be used to denote all pre-quenched operators and variables while 
not-primed ones for post-quench operators and variables. 
The initial state is then one of those in Eq. (\ref{ES}) for the pre-quench Hamiltonian, i.e. 
\be
|\Psi_0\rangle=|m_k\rangle=\prod_{k} ({b'}^\dag_k)^{m_k}|0\rangle. 
\label{psi0}
\ee 
As already discussed this state is fully specified by the characteristic function $m_k=0,1$ (in finite systems).  
The time dependent state is given by
\be
|\Psi_0(t)\rangle= e^{-iH_I t}|\Psi_0\rangle,
\ee
 with $H_I$ being the post-quench Hamiltonian with transverse field $h$.

We point out that even if the states $|m_k\rangle$ are a basis for the many-body Hilbert space, they 
do not represent the most generic excited state because the spectrum of the 
Ising chain is highly degenerate and linear combination of degenerate states are still eigenstates, but 
they cannot be written as (\ref{ES}).
One property of the states $|m_k\rangle$ is that they do not break the $Z_2$ symmetry 
of the Hamiltonian even in the ferromagnetic 
phase, unless $m_k=0$ i.e. in the ground state. 

The relation between the pre- and post-quench Bogoliubov operators is given by the combination of the two 
Bogoliubov rotations in Eq. (\ref{bogo_diag}) with angles $\theta_k$ and $-\theta'_k$ (i.e. post- and pre-quench Bogoliubov angles, respectively).
Thus the overall rotation is 
\be
B'_k=R_x(\theta_k-\theta'_k) B_k= R_x(\Delta_k) B_k,
\label{Rtot}
\ee
where we defined $\Delta_k\equiv\theta_k-\theta'_k$.
Being $\Delta_k$ the main quench variable entering in all the following calculations and results, 
it is worth writing explicitly its form in terms of $h$ and $h_0:$
\be
\cos \Delta_k=\frac{h h_0- (h+h_0) \cos \varphi_k+1}{\sqrt{1+h^2-2h\cos(\varphi_k)} \sqrt{1+h_0^2-2h_0\cos(\varphi_k)}}.
\ee

\subsection{Observables, Relaxation and Generalised Gibbs ensemble}
\label{sec.org}

The most important observable in the study of quantum quenches is the reduced density matrix $\rho_A$ of a 
block $A$ built with $\ell$ contiguous spins. Indeed from $\rho_A$ all the correlation functions local within $A$ can be obtained. 
Since the system is in a pure state $|\Psi_0(t)\rangle$ at any time, the density matrix of the entire system is
\be
\rho(t)=|\Psi_0(t)\rangle\langle\Psi_0(t)|.
\ee
The reduced density matrix of a subsystem $A$ is defined as
\be
\rho_A(t)={\rm Tr}_{\bar{A}}\big(\rho(t)\big),
\ee
where $\bar{A}$ is the complement of $A$.
The importance of $\rho_A$ stems from the fact that it is the quantity which generically displays a stationary behaviour described 
by some statistical ensemble, while the full density matrix $\rho(t)$ always corresponds to a pure state with zero entropy. 

More precisely, following Refs. \cite{bs-08,cef-ii} it is usually said that a system reaches a stationary state if a long time limit of the 
reduced density matrix exists, i.e. if the limit
\be
\lim_{t\to\infty}\rho_A(t) =\rho_A(\infty)
\ee
exists. 
This is described by a given statistical (mixed state) ensemble with full density matrix $\rho_E$
if its reduced density matrix restricted to $A$ equals $\rho_A(\infty)$, i.e. if 
for $\rho_{A, E}= {\rm Tr}_{\bar{A}}(\rho_E)$ and for any finite subsystem $A$
\be
\rho_A(\infty)=\rho_{A, E}\,.
\ee
In particular, this implies that arbitrary \emph{local} multi-point correlation functions within subsystem $A$ can be evaluated as
averages within the $\rho_E$.
By no means this implies that $\rho_E$ equals the full density matrix of the system which is clearly impossible being 
the former a mixed state and the latter a pure one. 

When a system thermalises, $\rho_E$ is the standard Gibbs distribution $\rho_E\propto e^{-\beta H}$ 
and this is expected to be the case when the model is non-integrable. 
However, for an integrable model, the proper statistical ensemble describing the system for long time 
is a generalised Gibbs ensemble (GGE)  rather than a thermal one. 
The density matrix of the GGE is defined as \cite{GGE}
\begin{equation}
\rho_{\text{GGE}}=\frac{e^{-\sum_n{\lambda_n I_n}}}{Z},
\label{GGE1}
\end{equation}
where $I_n$ is set of commuting integrals of motion, i.e. $[I_n,I_m]=0$, 
and $Z$ is a normalisation constant $Z=\mathrm{Tr} \,e^{-\sum_n{\lambda_n I_n}}$. 
It is important for a proper definition of GGE to specify which conserved charges enter in the GGE 
density matrix above. 
It has been understood  recently \cite{cef,cef-ii} that only {\it local} integrals of motion should be 
used in Eq. (\ref{GGE1}) if we are interested in the expectation values of {\it local} observables 
such as the reduced density matrix.

While in general it is a formidable task to calculate a reduced density matrix even for an integrable system, 
in the case of a model that can be mapped to free fermions it is a rather straightforward application of 
the Wick theorem to write it in terms of only two-point correlators of fermions. 
To this aim, let us first introduce the  \emph{correlation matrix} of Majorana fermions 
with the definition
\be
\braket{A_m A_n} = \delta_{mn} + i \Pi_{mn}\,.
\ee
For $\ell$ consecutive fermions ($2\ell$ Majoranas), using explicitly the periodicity of the chain, this matrix has the block structure
\begin{equation}\label{Ppi}
 \Pi =\left[
\begin{array}{cccc}
\Pi_0 & \Pi_{-1} & ... & \Pi_{1-\ell}\\
\Pi_1 & \Pi_0 & ... &... \\
... &... & ... & ...\\
\Pi_{\ell-1} & ... & ... & \Pi_0
\end{array}
\right],
\end{equation}
where the $\Pi_a$'s are $2\times 2$ matrices with entries equal to the correlations of Majorana fermions, which explicitly are
\be
\delta_{n0}-  i\Pi_n= 
\left(
\begin{array}{cc}
\braket{A_{2l-1} A_{2(l +n) -1} } & \braket{A_{2l-1} A_{2(l +n) } }\\
 \braket{A_{2l} A_{2(l +n) -1} }  & \braket{A_{2l} A_{2(l +n) } }
\end{array}
\right)
= 
\left(
\begin{array}{cc}
\braket{ A^y_l A^y_{l+n}} & \braket{A^y_l A^x_{l+n} }  \\
\braket{A^x_l A^y_{l+n} } & \braket{A^x_l A^x_{l+n} }
\end{array}
\right)\,,
\qquad \forall l,
\label{Pitop}
\ee
where the correlations can clearly be taken to start from an arbitrary site $l$. 
Because of its periodic structure, the matrix $\Pi$ turns out to be  a block Toeplitz matrix, i.e. its
constituent $2\times 2$ blocks depend only on the difference between row and column indices of $\Pi.$

We can now use Wick's theorem to construct all correlation functions
in the Ising chain.
As shown in Refs. \cite{vidal,pe-rev}, the matrix $\Pi$ determines entirely the 
reduced density matrix of the block $A$ of $\ell$ contiguous fermions in the chain
(and hence spins, because contiguous spins are mapped to contiguous
fermions, see Eq. (\ref{eq:sigmax})), with a final result that can be written in the compact way  \cite{vidal,fc-10}
\be
\rho_A=\frac{1}{2^\ell} \sum_{\mu_l=0,1}
\Bigl< \prod_{l=1}^{2\ell} A_l^{\mu_l}\Bigr>\left(\prod_{l=1}^{2\ell}
A_l^{\mu_l}\right)^\dag \propto e^{A_l W_{lm} A_m/4}\,, 
\label{quad}
\ee
where  \cite{fc-10}
\be
\tanh\frac{W}2=i \Pi\, . 
\ee
Given $\rho_A$ we can calculate any local correlation function with support in $A$. 
Very importantly, because of this direct relation between reduced density matrix and correlation matrix, 
it is sufficient to prove that the two ensembles or states have the same correlation matrix in order to 
prove that they are equal. This is an immense simplification because while $\rho_A$ has $2^\ell\times 2^\ell$
elements, $\Pi$ has only $2\ell\times 2\ell$ elements. 

From the reduced density matrix, another fundamental observable is easily constructed, namely the entanglement 
entropy which is the von Neumann entropy of $\rho_A$,
\be
S_A=- {\rm Tr} \rho_A \ln \rho_A\,.
\ee
Using again Wick's theorem \cite{vidal}, $S_A$ can be related to the eigenvalues of the matrix $\Pi$. 
Indeed, denoting the eigenvalues of  $\Pi$ as $\pm i\nu_m$, $m=1...\ell$ (being $\Pi$ an antisymmetric  matrix, its 
eigenvalues are purely imaginary complex conjugate pairs), 
the entanglement entropy is \cite{vidal}
\be S=\sum_{m=1}^\ell H(\nu_m),\qquad {\rm where}\quad
H(x)=-\frac{1+x}{2} \ln\left(\frac{1+x}{2}\right)-\frac{1-x}{2}\ln\left(\frac{1-x}{2}\right).
\label{SvsH}
\ee

Apart from the reduced density matrix, we will also consider the transverse magnetisation
\begin{equation}\label{m_z}
m^z(t)=\braket{\sigma^z_i}=\langle i A^x_i A^y_i \rangle,
\end{equation}
and the two-point function of the order parameter at a distance $\ell$ 
\begin{equation}
\rho^{xx}(\ell)\equiv \langle \sigma^x_n \sigma^x_{\ell+n}\rangle . 
\end{equation}
While $m^z$ is local within fermions, $\rho^{xx}(\ell)$ is not. 
However, in computing $\rho^{xx}(\ell)$ only the string of Majorana fermions between sites 
$n$ and $n+\ell$ matters and thus it takes the form \cite{bm-70,sach-book} 
\begin{equation}
\rho^{xx}(\ell)=\left\langle\prod_{j=n}^{\ell+n-1}(-i A^y_j A^x_{j+1})\right\rangle.
\end{equation}
By means of Wick's theorem \cite{bm-70,sach-book}, $\rho^{xx}(\ell)$ can be written  
as the Pfaffian of a skew symmetric $2\ell\times 2\ell$ matrix
\begin{equation}
\rho^{xx}(\ell)=\mathrm{pf}(\Gamma),
\end{equation}
where 
$\Gamma$ is given by
\begin{equation}\label{gamma}
\Gamma=\left[
\begin{array}{cccc}
\Gamma_0 & \Gamma_{-1} & ... & \Gamma_{1-\ell}\\
\Gamma_1 & \Gamma_0 & ... &... \\
... &... & ... & ...\\
\Gamma_{\ell-1} & ... & ... & \Gamma_0
\end{array}
\right],
\end{equation}
where
\be
\delta_{n0}-  i\Gamma_n= 
\left(
\begin{array}{cc}
\braket{ A^y_l A^y_{l+n}} & \braket{A^x_l A^y_{l+n-1} }  \\
\braket{A^y_l A^x_{l+n+1} } & \braket{A^x_l A^x_{l+n} }
\end{array}
\right)\,,
\qquad \forall l.
\label{gammael}
\ee
Notice that although the matrices $\Gamma$ and $\Pi$ in Eq. (\ref{Pitop}) look very similar and they have the 
same block-diagonal elements, the off-diagonal ones are different since the second operator is shifted by $\pm1$.
We note that in the literature the two matrices are often denoted by the same symbol and it is very easy to mix them up.

\section{The infinite time limit and the generalised Gibbs ensemble}
\label{equivalence}

In this section we consider the infinite time limit of the reduced density
matrix $\rho_A$ of a subsystem A composed of $\ell$ contiguous spins. 
To analyse $\rho_A$, we first consider the time evolution of the long time limit of its building
blocks, that, according to Eq. (\ref{quad}), are the two-point real-space correlation functions of fermions.

\subsection{Time evolution of the fermionic two-point function} 

The Bogoliubov rotations diagonalising the pre- and post-quench Hamiltonians  only couple modes with 
opposite momenta, cf. Eq. (\ref{bog2}).
It is then convenient to cast the two-point correlation functions of pre-quench Bogoliubov modes in the $2\times 2$ matrix 
\begin{equation}
\begin{split}
\langle\Psi_0| B'_k B_k^{'\dagger}|\Psi_0\rangle &=\left(
\begin{array}{cc}
\langle b_k'b_{k}^{'\dagger}\rangle & \langle b_k'b'_{-k}\rangle \\
\langle b_{-k}^{'\dagger}b_{k}^{'\dagger}\rangle & \langle b_{-k}^{'\dagger} b'_{-k}\rangle
\end{array}
\right)=\left(
\begin{array}{cc}
1-m_k & 0 \\
0 & m_{-k}
\end{array}
\right)\\ &=\frac{1}{2}\left[\sigma_z(1-m_k-m_{-k})+{\mathbb I}(1-m_k+m_{-k})\right],
\end{split}
\end{equation}
in which $|\Psi_0\rangle$ is the initial state specified by the function $m_k$ as in Eq. (\ref{psi0}). 
Combining the two Bogoliubov rotations for pre- and post-quench Hamiltonian as in Eq. (\ref{Rtot}),
we write  the expectation value of post-quench Bogoliubov operators in the initial state as
\begin{equation}\label{BB}
\begin{split}
\langle\Psi_0| B_k B_k^{\dagger}|\Psi_0\rangle & =\langle\Psi_0|R_x(-\Delta_k) B'_k B_k^{'\dagger}R_x^\dagger(-\Delta_k)|\Psi_0\rangle\\
&=\left(
\begin{array}{cc}
 \sin^2{\frac{\Delta_k}{2}} m_{-k}+\cos^2{ \frac{\Delta_k}{2}}(1-m_k) & -\frac{i}{2}\sin{\Delta_k} (-1+m_{-k}+m_k) \\
 \frac{i}{2}  \sin{\Delta_k} (-1+m_{-k}+m_k) & \cos^2{\frac{\Delta_k}{2}} m_{-k}+\sin^2{\frac{\Delta_k}{2}} (1-m_k)
\end{array}
\right),
\end{split}
\end{equation}
which is the initial condition for the fermionic two-point functions. 
The time evolution can be worked out in the Heisenberg picture where the 
post-quench operators $B_k(t)$ evolve according to the Hamiltonian (\ref{H_diag_preq}), 
so $B_k(t)=U_k(t)B_k(0)$, where $U_k(t)$ is the restriction of the time evolution operator to the 
subset of the Hilbert space with momenta $k$ and $-k$, i.e. 
\begin{equation}\label{evo}
U_k(t)
=\left(
\begin{array}{cc}
e^{-i\epsilon_k t} & 0 \\
0 & e^{i\epsilon_k t}
\end{array}
\right)=R_z(-2\epsilon_k t).
\end{equation}

It is now possible to evaluate the expectation value of expressions bilinear in the
fermions $c_i$ and $c_i^\dagger$ at any time. 
In order to do so, it is enough to 
invert Eq. (\ref{def_d}) and express the $d$ operators in terms of $b$ which evolve according to Eq. (\ref{evo}).
From $c_l= \sum_k e^{i2\pi kl/N}(u_k b_k+iv_k b^\dagger_{-k})/\sqrt{N}$, we can write
\begin{equation}
c_l(t)=\frac{1}{\sqrt N}\sum_k e^{i2\pi kl/N}\left(u_k b_k(t)+iv_k b^\dagger_{-k}(t)\right)=\frac{1}{\sqrt N}
\sum_k e^{i2\pi kl/N}\left(u_k e^{-i\epsilon_k t}b_k+iv_k e^{i\epsilon_k t}b^\dagger_{-k}\right),
\end{equation}
and similarly for $c_l^\dagger(t)$.
Hence from Eq. (\ref{Omega}) we obtain for the Fourier transform of Majorana operators 
\begin{equation}\label{omega_k}
\omega_k^+(t)=e^{i\theta_k/2}\left(b_k e^{-i\epsilon_k t}+b_{-k}^\dagger e^{i\epsilon_k t}\right),\quad\quad \omega_k^-(t)=
e^{-i\theta_k/2}\left(b_{-k}^\dagger e^{i\epsilon_k t}-b_{k} e^{-i\epsilon_k t}\right) .
\end{equation}
The vector operator $\Omega_k(t)$ has the form
\begin{equation}
\Omega_k(t)=\left(
\begin{array}{cc}
-e^{-i(\theta_k/2+\epsilon_k t)} & e^{i(-\theta_k/2+\epsilon_k t)}  \\
e^{i(\theta_k/2-\epsilon_k t)} & e^{i(\theta_k/2+\epsilon_k t)}
\end{array}
\right)B_k.
\end{equation}
Thus
\begin{equation}\label{om_t}
\begin{split}
&\langle\Omega_k(t)\Omega_k^\dagger(t)\rangle =\langle\Psi_0|\left(
\begin{array}{cc}
-\omega_k^-(t)\omega_{-k}^-(t) & \omega_k^-(t)\omega_{-k}^+(t)  \\
-\omega_k^+(t)\omega_{-k}^-(t)  &\omega_k^+(t)\omega_{-k}^+(t) 
\end{array}
\right)|\Psi_0\rangle\\ &=\left(
\begin{array}{cc}
1+m^A_k-m^S_k\sin{2t\epsilon_k}\sin{\Delta_k} & m^S_k e^{-i\theta_k}(\cos{\Delta_k}-i\cos{2t\epsilon_k}\sin{\Delta_k}) \\
m^S_k e^{i\theta_k}(\cos{\Delta_k}+i\cos{2t\epsilon_k}\sin{\Delta_k})  &1+ m^A_k +m^S_k\sin{2t\epsilon_k}\sin{\Delta_k}
\end{array}
\right),
\end{split}
\end{equation}
where we defined 
\bea
m^S_k&\equiv& m_{-k}+m_{k}-1,\\ 
m^A_k&\equiv& m_{-k}-m_{k},
\eea
which stand for the even and odd part of $m_k$ respectively. 
It is straightforward to check that when $m_k$ and $m_{-k}$ are set to zero this expression reduces to the one 
obtained in the case when the initial state is the ground state of the initial Hamiltonian \cite{cc-05,cef-ii}.
Note that if $m^S_k=0$ then Eq. (\ref{om_t}) is constant in time, which is a manifestation of the fact that 
if $m^S_k=0$ for any $k$ the state is not only an eigenstate of the pre-quench Hamiltonian but also of the post-quench one.
We will define the states with $m^A_k=0$ as {\it parity invariant states} (PIS) in which all the positive and negative momentum modes 
are populated with the same weights. Note that all the PIS have zero momentum, but the condition for PIS is 
more restrictive than that. 
We will refer to the states with $m^A_k\neq 0$ for some $k$ as {\it non parity invariant states} (NPIS).

Equation (\ref{om_t}) is the final expression for the two-point function of fermions in momentum space from which, by Fourier 
transform, the one for real-space fermions given by Eq. (\ref{Pitop}) can be straightforwardly obtained. 
As a usual feature  of free systems, each momentum mode oscillates in time with typical frequency 
proportional to $\epsilon_k$. However, when taking the Fourier transform, in the {\it thermodynamic limit}
the various modes interfere in a destructive way and their long-time expectation is the time-average of the expression above, i.e. 
\begin{equation}\label{time_av_Om}
\langle\overline{\Omega_k(t)\Omega_k^\dagger(t)}\rangle=\left(
\begin{array}{cc}
1+m^A_k & m^S_k e^{-i\theta_k}\cos{\Delta_k}  \\
m^S_k e^{i\theta_k}\cos{\Delta_k} & 1+ m^A_k,
\end{array}
\right).
\end{equation}

Thus, in order to show that the reduced density matrix attains a stationary behaviour described by GGE, 
it is sufficient to show that the GGE prediction for  $\langle\Omega_k\Omega_k^\dagger\rangle$
equals Eq. (\ref{time_av_Om}). 
By no means this implies that $\langle\Omega_k(t)\Omega_k^\dagger(t)\rangle$ has a long-time limit, 
on the contrary, it oscillates forever as a consequence of the fact that the state is pure for any time and the Hamiltonian governing the evolution is diagonal in the modes.

\subsection{GGE expectation value of the fermionic two-point function}

The GGE density matrix for the full system is given by Eq. (\ref{GGE1}) constructed with the local integrals of motion. 
However, it has been shown that for the transverse field Ising chain the post-quench 
occupation number operators 
\be 
n_k=b_k^\dagger b_k,
\ee
although non-local quantities, can be written as linear combinations of the local integrals of motion \cite{fe-13a}
(see also the next subsection). 
Thus the GGE density matrix constructed with local integrals of motion and the one constructed with 
$n_k$ are equivalent. 
We will then consider
\begin{equation}
\rho_{\rm GGE}=\frac{e^{-\sum_k{\lambda_k n_k}}}{Z},
\end{equation}
where we use the same symbols for the  Lagrange multipliers $\lambda_k$ and $\lambda_n$ 
in Eq. (\ref{GGE1}) since we will never use the two concomitantly. 
The $\lambda_k$ are  fixed by matching the expectation values of the occupation numbers with their values
in the initial state, i.e. imposing 
\begin{equation}
\langle\Psi_0| b_k^\dagger b_k|\Psi_0\rangle=\mathrm{Tr}[\rho_{\text{GGE}}\, b_k^\dagger b_k].
\end{equation}
The left-hand-side of this equation can be read out from Eq. (\ref{BB})
\begin{equation}\label{n_k}
\langle n_k\rangle= \langle\Psi_0| b_k^\dagger b_k|\Psi_0\rangle= 
1-\sin^2\left(\frac{\Delta_k}{2}\right)m_{-k}-\cos^2\left(\frac{\Delta_k}{2}\right)(1-m_k),
\end{equation}
while the right-hand-side is
\begin{equation}
\braket{n_k}= Tr[\rho_{\text{GGE}} \,b_k^\dagger b_k]=\frac{1}{1+e^{\lambda_k}}.
\end{equation}
Equating the two expressions we get the equation determining  $\lambda_k$:
\begin{equation}
\frac{1}{1+e^{\lambda_k}}=1-\sin^2\left(\frac{\Delta_k}{2}\right)m_{-k}-\cos^2\left(\frac{\Delta_k}{2}\right)(1-m_k).
\end{equation}
The components of $\langle\Omega_k\Omega_k^\dagger\rangle$ can be readily calculated in the GGE, for example 
\bea
\langle\omega_k^+\omega_{-k}^+\rangle&=&\mathrm{Tr}[\rho_{\text{GGE}}\,\omega_k^+\omega_{-k}^+]=
\frac{1}{Z}\mathrm{Tr}[e^{-\sum_k{\lambda_k b^\dag_kb_k}}(1-b^\dag_kb_k+b^\dag_{-k}b_{-k})]\\
&=&1-\frac{1}{1+e^{\lambda_k}}+\frac{1}{1+e^{\lambda_{-k}}} 
=1-\braket{n_k}+ \braket{n_{-k}}=1+m_{-k}-m_k=1+ m_k^A\,. \nonumber
\eea
Performing  similar calculations for the other three elements of the matrix $\langle\Omega_k\Omega_k^\dagger\rangle$ we finally get
\be
\langle\Omega_k\Omega_k^\dagger\rangle_{\rm GGE}=\left(
\begin{array}{cc}\label{omega(t)}
1+ m^A_k & m^S_k e^{-i\theta_k}\cos{\Delta_k}  \\
m^S_k e^{i\theta_k}\cos{\Delta_k}  & 1+ m^A_k
\end{array}
\right),
\ee
which coincides with Eq. (\ref{time_av_Om}). 
This proves that the GGE two-point functions of fermions at arbitrary distance are equal to
the long-time limit of the same two-point function after a quench from an excited state $|m_k\rangle$
of the initial Hamiltonian. 
Since the reduced density matrix can be constructed solely from the fermionic two-point functions 
as in Eq. (\ref{quad}), this also proves that any local multipoint correlation function of spins and fermions
will be described by the GGE for long times.

\subsection{Local Conservation laws in the TFIC}
\label{laws}

It is instructive to have a look at  the behaviour of the local conservation laws in the chain in order to check
whether they bring any further understanding in the non-equilibrium quench dynamics.  
In the thermodynamic limit  there is an infinite number of local conserved charges 
which can be written in terms of the post-quench occupation numbers as \cite{p-98,fe-13a}
\begin{alignat}{2}
\label{I+}
I_n^-&=-&&\int_{-\pi}^{+\pi}\frac{dk}{2\pi}\sin[(n+1)k]b_k^\dagger b_k,\\
I_n^+&=&&\int_{-\pi}^{+\pi}\frac{dk}{2\pi}\cos(nk)\epsilon_k b_k^\dagger b_k,\quad \quad n\geq0\nonumber,
\end{alignat}
where the apex $\pm$ refers to their parity properties: $I_n^+$ are even and $I_n^-$ are odd under spatial reflections. 
The expectation values of these conserved charges in the initial state (and in the subsequent time evolution) is 
obtained by inserting the expectation value of $n_k$ of Eq. (\ref{n_k}) into Eq. (\ref{I+}):
\begin{alignat}{2}
\label{I_n_sempli}
\braket{I_n^-} &=-&&\int_{-\pi}^{+\pi}\frac{dk}{4\pi}\sin[(n+1)k]m_k^A,\\
\braket{I_n^+}&=&&\int_{-\pi}^{+\pi}\frac{dk}{4\pi}\cos(nk)\epsilon_k\left[1+m_k^S\cos\Delta_k\right]. \nonumber
\end{alignat}
Hence $\braket{I_n^+}$ and $\braket{I_n^-}$ are both, in general, non-vanishing for an initial excited state. 
This is different from what happens 
if the initial state is the ground state of the pre-quench Hamiltonian, when all the parity-odd charges $I_n^-$ vanish.
However, $\braket{I_n^-}$ is zero every time that $m_k^A=0$, i.e. for PIS with $m_k=m_{-k}$. 
This will represent an important class of states for which the calculation of some local observables in the following 
will be much easier. 
It is surely interesting to understand  whether the increased number of non-zero conservation laws alters somehow
the time-dependence of the asymptotic behaviour of local correlation functions. 

Finally  we would like to emphasise that the $\braket{I_n^-}$'s depend only on the initial state and not
on the quench parameters, which are entirely contained in the $\Delta_k$ angle that instead appears in 
$\braket{I_n^+}$. The independence of $\langle I^-_n\rangle$ on the Bogoliubov transformation was also pointed out in Refs. \cite{fe-13a,f-13}.

\section{Transverse magnetisation}
\label{magne}

 The first observable we consider is the transverse magnetisation which is particularly easy to calculate because 
it has a local expression in terms of fermions, cf. Eq. (\ref{m_z}).

In the case when the initial state is the pre-quench ground state the transverse magnetisation 
is in the thermodynamic limit  \cite{bm-70,cef-ii}
\begin{equation}\label{m_ground}
m^z(t)=-\int_{-\pi}^\pi\frac{dk}{4\pi}e^{i\theta_k}[\cos\Delta_k-i\sin\Delta_k\cos(2\epsilon_k t)].
\end{equation}
For an excited state with characteristic function $m_k$, $m^z(t)$  
can be easily found expressing the $c_i$ in terms of the $\omega_k^+(t), \omega_k^-(t)$ 
and substituting the matrix elements of Eq. (\ref{om_t}). 
After simple algebra we obtain in the thermodynamic limit
\begin{equation} \label{mz_ex}
m^z(t)=\int_{-\pi}^\pi\frac{dk}{4\pi}e^{i\theta_k}m^S_k[\cos\Delta_k-i\sin\Delta_k\cos(2\epsilon_k t)],
\end{equation}
which again reduces to the ground-state evolution in the case $m^S_k=-1$, i.e. $m_k=0$. 
Quite remarkably, we have that only the symmetric part of the characteristic function $m_k$ contributes
to the time evolution of the transverse magnetisation and so the odd conserved charges in Eq. (\ref{I_n_sempli}) 
have no influence at all on this observable. 

Furthermore, the result can be divided into a stationary and a time-dependent part. 
As for the ground-state case, it is particularly interesting to understand the approach to the stationary value 
which can be evaluated by a stationary phase approximation. 
The stationary points are the zeros of $\epsilon'_k=d \epsilon_k/dk$ in the interval $[-\pi,\pi]$,  which are $-\pi, 0, \pi$.  
However, when the characteristic function $m_k$ is not an analytic function in $k$, 
possible new extrema of the integration domain need to be taken into account in the calculation. 
Indeed, these extrema need not coincide with $-\pi, \pi$, as it happens for example in the case in which the initial state 
excitations $m_k$ are non-zero only in a particular subinterval of $[-\pi, \pi]$.

In the general case under consideration, we have integrals of the form
\begin{equation}
I(t)=\int_a^b dk f(k) e^{i t g(k)},
\label{Iab}
\end{equation}
where $a$ and $b$ do not need to coincide with $-\pi$ and $\pi$. 
Notice also that in Eq. (\ref{mz_ex})  the stationary points of $g(k)$ are always zeros of $f(k)$, requiring to go to the second
order in stationary phase approach. 
There are three different classes of extremal points which should be 
summed up in order to have the complete large time behaviour. 
Denoting with $k_0$ each of the points, the three classes are
\begin{itemize}
\item[1)] The extremal point $k_0$ is internal to the domain of integration, i.e. $k_0\in (a,b)$. In this case $k_0$ must be 
stationary to be extremal, i.e. $\epsilon'(k_0)=0$, and then we have 
\begin{equation}
I(t)=I_{0}(t)\equiv e^{i t g(k_0)} \left[\frac{f''(k_0)}{2}\frac{\sqrt\pi}{2}\frac{1}{(-a')^{3/2}}+O[(-a')^{-5/2}]\right],
\label{stat1}
\end{equation}
where $a'=\frac{i t}{2} g''(k_0)$. Because of $f(k_0)=0$ the leading behaviour is $(-a')^{-3/2}$ instead of $(-a')^{-1/2}$.
\item[2)] The extremal point $k_0$ is at the boundary of the integration domain, i.e. $k_0= a$ or  $b$ and 
it is also stationary, i.e. $\epsilon'(k_0)=0$, in which case we have
\begin{equation}
I(t)=\frac{1}{2} I_{0} (t).
\label{stat2}
\end{equation}
\item[3)] Finally, the extremal point can be at the boundary of the integration domain, i.e. $k_0= a$ or  $b$,
but it is not stationary i.e. $\epsilon'(k_0)\neq0$. For $k_0= a $ and not stationary we have 
\begin{equation}
I(t)=-f(a)\frac{e^{i t g(a)}}{i t g'(a)},
\end{equation}
and the same without the minus sign for $k_0= b$.
\end{itemize}

It is clear that cases (1) and (2) give a different contribution compared to case (3). 
Indeed the first two cases generically give a large time behaviour of the type  $t^{-3/2}$ (whenever $f''(k_0)\neq 0$, else 
the contribution would be faster like $t^{-5/2}$), like in the quench from the ground state. 
The third case instead will generically produce a slower decay of the type $t^{-1}$ (unless $f(a)=0$, in which 
case we will have $t^{-2}$). 

For a generic initial state with characteristic function $m_k$, the general strategy would be the following:
(i) divide the integral in Eq. (\ref{mz_ex}) in pieces in which $m^S_k$ is continuous, (ii) treat each of the integrals
as in Eq. (\ref{Iab}), and finally (iii) sum up all the contributions with the slowest power-law behaviour. 
In order to show the differences between the various states, we compare the evolution from
the ground state with the ones from three representative excited states.
We choose the following three states
\begin{align}
m_1(k)&=\theta(k-\pi/2),\nonumber\\
m_2(k)&=(k/\pi)^2,\nonumber\\
m_3(k)&=(k+\pi)/(4\pi),
\end{align}
where all $m_a(k)$ are clearly defined in the interval $[-\pi,\pi]$.

For the initial ground state we regain the large time behaviour \cite{cef-ii}
\begin{equation}\label{timedep}
m^z_{0}(t)\simeq m^z_{\rm stat0}-\frac{c(t)}{(hJt)^{3/2}},
\end{equation}
where 
\begin{equation}
c(t)=\frac{(h-h_0)}{32\sqrt{2\pi}J}\left[ \frac{\cos(\pi/4 + 4Jt(1+h))}{(1+h_0)\sqrt{1+h}}+\frac{\sin(\pi/4 + 4Jt |h-1|)}{|h_0-1|\sqrt{|h-1|}}   \right].
\end{equation}
For the initial state characterised by $m_1(k)$ we obtain
\begin{equation}
m^z_{1}(t)\simeq m^z_{\rm stat1}- \frac{c_{11}(t)}{(hJt)}- \frac{c_{12}(t)}{(hJt)^{3/2}},
\label{statm1}
\end{equation}
where 
\begin{align}
c_{11}(t)&=-\frac{(h-h_0)}{16J\pi} \frac{\sin( 4Jt\sqrt{1+h^2})}{\sqrt{(1+h^2)(1+h_0^2)}},\nonumber\\
c_{12}(t)&=\frac{(h-h_0)}{64J\sqrt{\pi}} \frac{[\sin( 4Jt|h-1|)+\cos{4Jt(h-1)}]}{\sqrt{|h-1|}|h_0-1|}.
\end{align}
The leading large time behaviour going like $t^{-1}$ comes from the extrema $\pm \pi/2$ (non-stationary points) 
while the $t^{-3/2}$ from the $k=0$ stationary point.
Conversely, for the initial states characterised by $m_2(k)$ and $m_3(k)$ we get a $t^{-3/2}$ power law behaviour 
due to the stationary points $x=0,\pm\pi$ falling into cases $1)$ and $2)$ above,
similarly to the ground state but with different calculable coefficients that we do not report here, but 
are easily obtained from Eqs. (\ref{stat1}) and (\ref{stat2}).

\begin{figure}[t] 
\begin{center} 
\includegraphics[width=.49\textwidth]{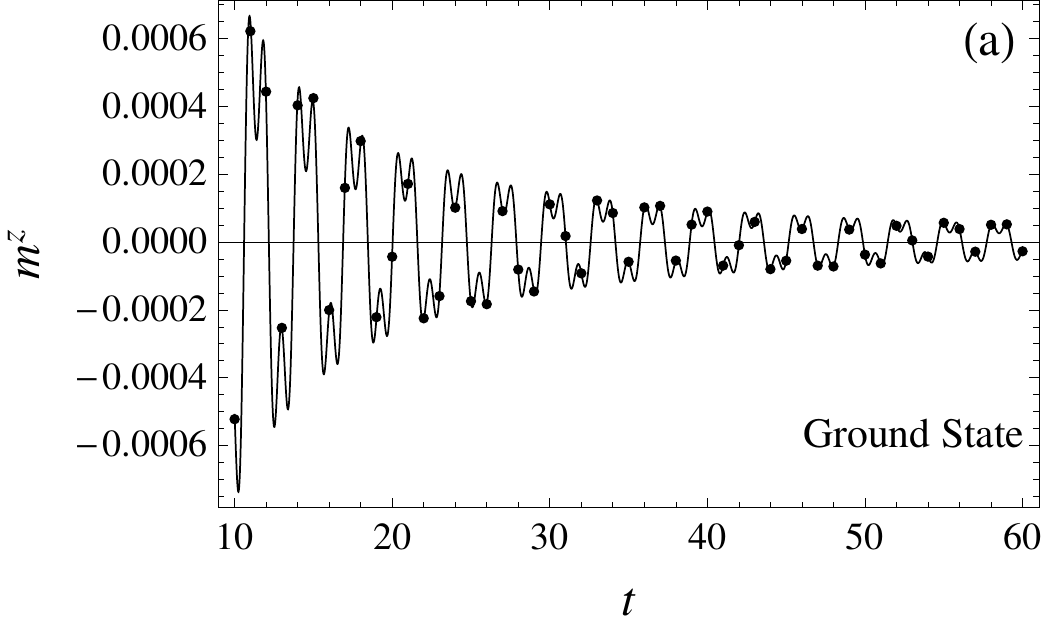}~\includegraphics[width=.49\textwidth]{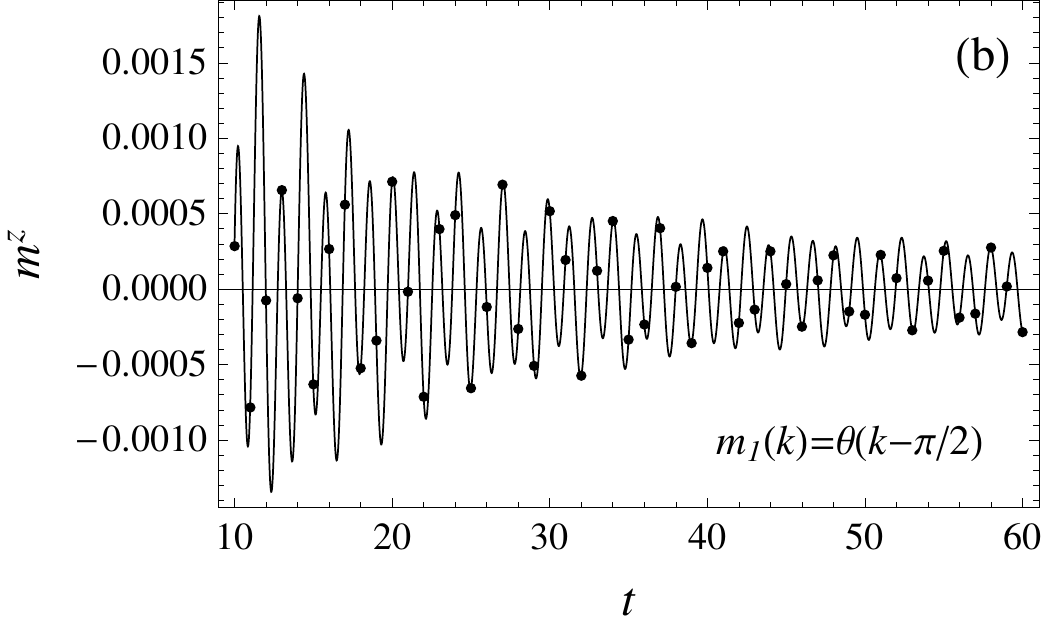}\\
\includegraphics[width=.49\textwidth]{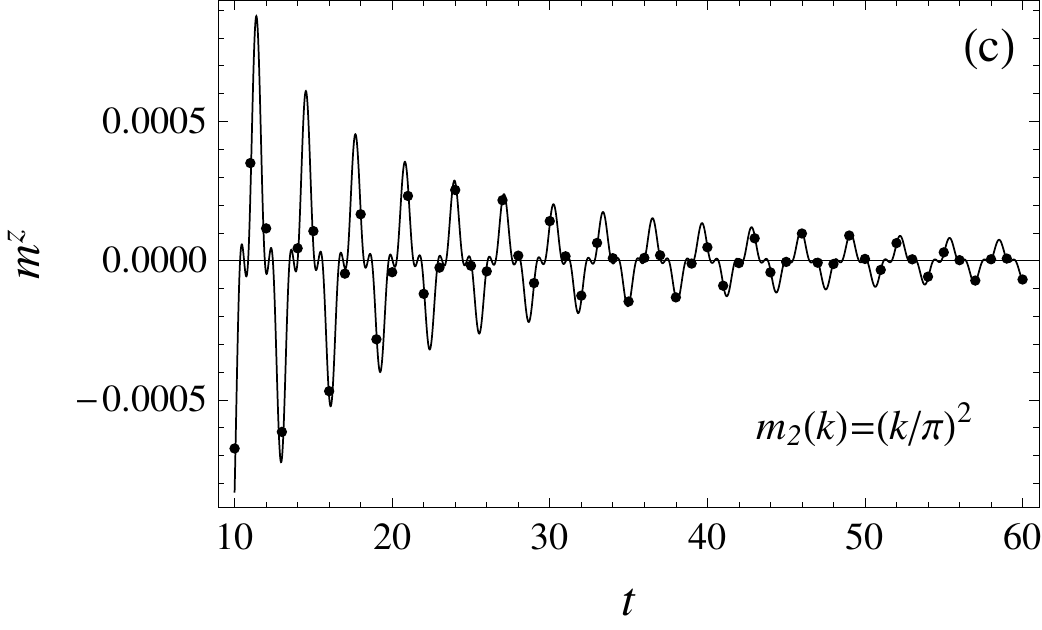}~\includegraphics[width=.49\textwidth]{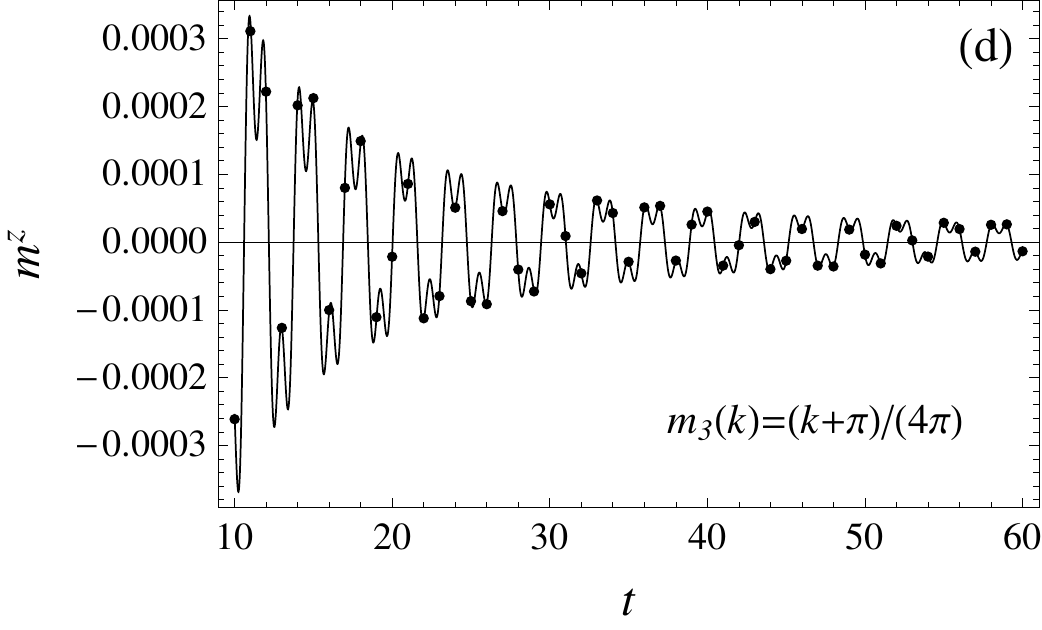}
\caption{\label{fig:graficimagne}\small Transverse magnetisation minus its stationary value as a function of time for quenches from $h_0=12$ to $h=2$ for different initial states.
The points represent the exact evaluation of the integral in Eq. (\ref{mz_ex}) while the lines 
are the stationary phase approximation results valid for large time.
a) Ground state. b) $m_1(k)=\theta(k-{\pi}/{2})$.   c) $m_2(k)=({k}/{\pi})^2$. d) $m_3(k)=({k+\pi})/({4\pi})$.
In all the cases the agreement is excellent for large enough time.}
\end{center} 
\end{figure} 

In order to show the reliability and the range of validity of the stationary phase approximations,
in Fig. \ref{fig:graficimagne} we report the time dependent part of the transverse magnetisation
(i.e. we subtract the stationary behaviour). We compare the exact results from the numerical 
determination of the integral in Eq. (\ref{mz_ex}) with the stationary phase approximation 
up to order $t^{-3/2}$. 
It is evident that even for not so-large time the oscillating power-law decay of the
stationary phase correctly describes the data. 
We mention that in the case of Eq. (\ref{statm1}) it is important to keep the term $t^{-3/2}$ to describe the data for not too 
large times. 

To conclude this section we would like to emphasise the main difference we have found 
in the large time behaviour of the transverse magnetisation starting from the ground-state 
or from excited states of the pre-quench Hamiltonian. 
For quenches starting from the ground-state we always have a power-law tail of the form $t^{-3/2}$. 
While several excited states have the same power-law behaviour, this is not true in general.
The state with $m_1(k)$ above presents a much slower relaxation going like $t^{-1}$.
We stress that this is not at all an academic state, quite the reverse, it is the most physical among the ones presented above 
because it has all the modes larger than a given one ($\pi/2$, but this is not essential) occupied. 
Furthermore by choosing very particular momentum occupation functions $m(k)$, it is not difficult to 
cook up quite untypical power-law behaviours: for example, considering $m(k)=\sin^2(k)$, since it vanishes in all the stationary points of the phase $\epsilon(k)$, we  obtain a power-law decay like  $t^{-5/2}$.

\section{Equal-time two point longitudinal correlation function}
\label{longe}

In this section we investigate the longitudinal spin-spin correlation function between two spins at a distance 
$\ell$ at the same time $t$, i.e. 
\begin{equation}
\rho^{xx}(\ell,t)\equiv \langle \Psi_0(t)| \sigma^x_n \sigma^x_{\ell+n}|\Psi_0(t)\rangle . 
\end{equation}
The most interesting regime of this two-point function is the so-called {\it space-time scaling limit} \cite{cef,cef-i}
defined as the limit $t,\ell\to\infty$ with their ratio $t/\ell$ kept fixed.
In general, the space-time scaling limit does not have to commute with the limit $t\to 0$ or $t\to\infty$ either if taken before or after the thermodynamic limit.

The two-point function $\rho^{xx}(\ell,t)$ is the Pfaffian of the $2\ell\times 2\ell$ matrix $\Gamma$ in Eq. (\ref{gamma}) the elements of which are the already calculated two-point fermion functions in Eq. (\ref{gammael}).
The fermionic correlators in Eq. (\ref{gammael}) can be identified looking at Eq. (\ref{om_t}), obtaining 
that the two-by-two constituent blocks have the form 
\be
\Gamma_n=\left(
\begin{array}{cc}
h_n & g_n\\
-g_{-n} & f_n
\end{array}
\right),
\ee
with elements
\begin{align}
f_n+i\delta_{n0}&\equiv  i\langle A^x_j A^x_{j+n}\rangle 
=\frac{i}{N}\sum_k e^{i2\pi kn/N}\left[1+m^A_k-m^S_k\sin{2t\epsilon_k}\sin{\Delta_k}\right],
\label{fdef}\\
h_n+i\delta_{n0}&\equiv  i\langle A^y_{j} A^y_{j+n}\rangle 
=\frac{i}{N}\sum_k e^{i2\pi kn/N}\left[1+ m^A_k+m^S_k\sin{2t\epsilon_k}\sin{\Delta_k}\right],
\label{hdef}\\
g_n&\equiv  i \langle A^x_j A^y_{j+n-1}\rangle 
=\frac{1}{N}\sum_k e^{i2\pi kn/N}\left[-m^S_k e^{-ik}e^{i \theta(k)}(\cos{\Delta_k}-i\cos{2t\epsilon_k}\sin{\Delta_k})\right]\label{gdef}.
\end{align}
Thus in the thermodynamic limit we have 
\begin{equation}
\label{Gamma_k}
\Gamma_n=\left(
\begin{array}{cc}
h_n & g_n\\
-g_{-n} & f_n
\end{array}
\right)=\int_{-\pi}^{\pi}\frac{dk}{2\pi}e^{ikn}\hat\Gamma(k)\,, \quad \mathrm{with}\quad 
\hat\Gamma(k)=\left(
\begin{array}{cc}
h(k) & g(k)\\
-g(-k) & f(k)
\end{array}
\right),
\end{equation}
with
\begin{align}
f(k)&= i\left[m^A_k-m^S_k\sin{2t\epsilon_k}\sin{\Delta_k}\right],\nonumber\\
h(k)&= i\left[m^A_k+m^S_k\sin{2t\epsilon_k}\sin{\Delta_k}\right],\nonumber\\
g(k)&= -m^S_k e^{-ik}e^{i \theta(k)}\left(\cos{\Delta_k}-i\cos{2t\epsilon_k}\sin{\Delta_k}\right) .
\end{align}
The  $2\times2$ matrix $\hat\Gamma(k)$ is called the {\it block symbol} of the matrix $\Gamma$.
Since $f(k)$ and $h(k)$ are odd functions of $k$, the matrix ${\Gamma}$ is antisymmetric and of  Toeplitz form, as it should be.
Notice that when the initial state is the ground state of the pre-quench Hamiltonian $h(k)=-f(k)$, which is 
also the case every time when $m^A_k=0$, i.e. $m_k=m_{-k}$.

\subsection{Numerical results}

\begin{figure}[t!] 
\begin{center} 
\includegraphics[width=.49\textwidth]{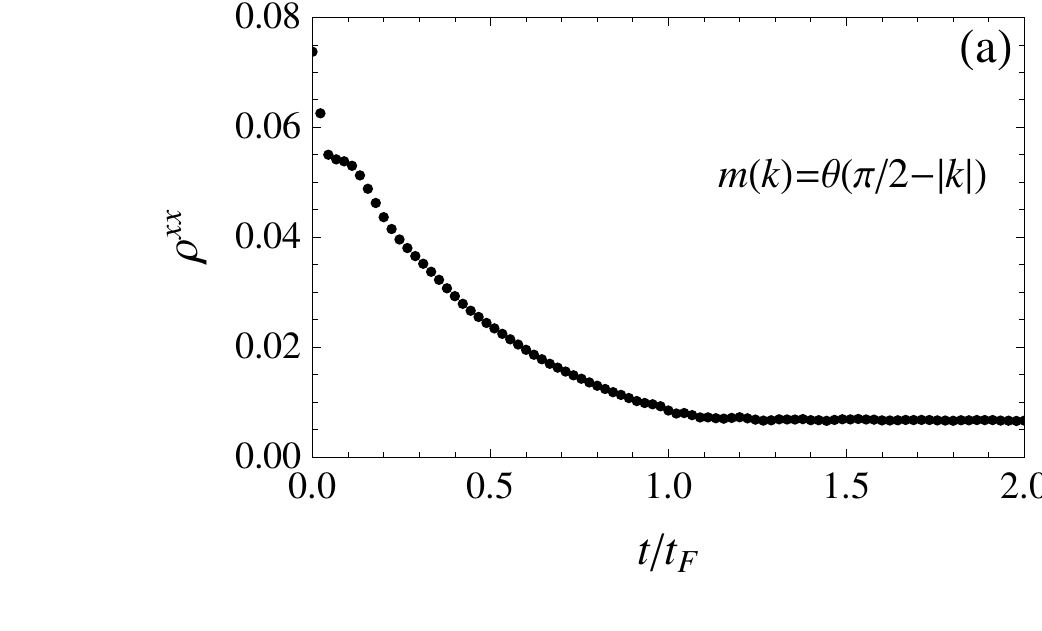}~\includegraphics[width=.49\textwidth]{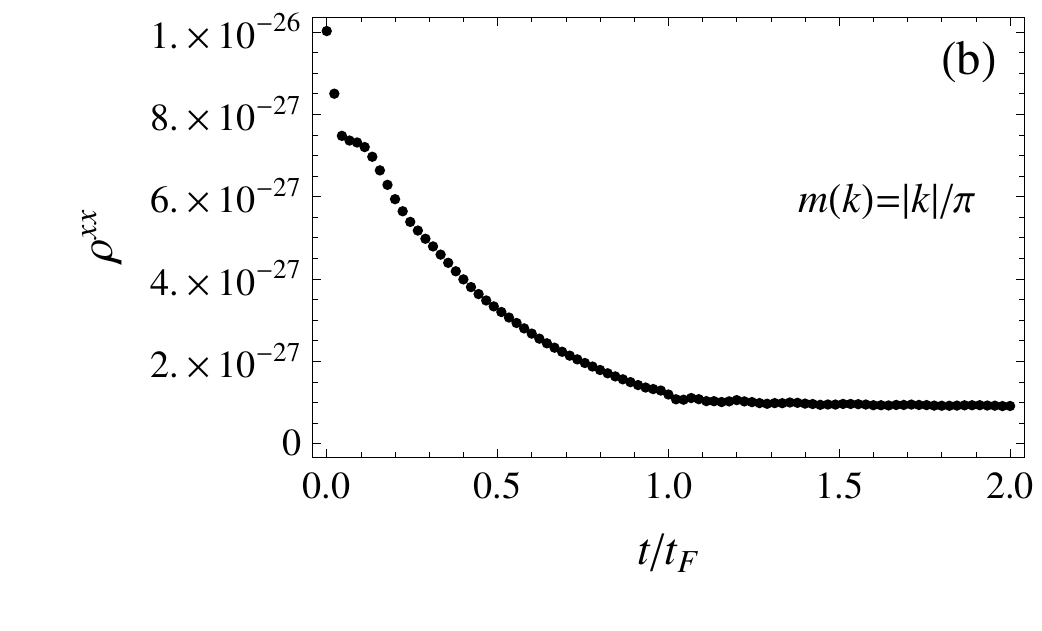}\\
\includegraphics[width=.49\textwidth]{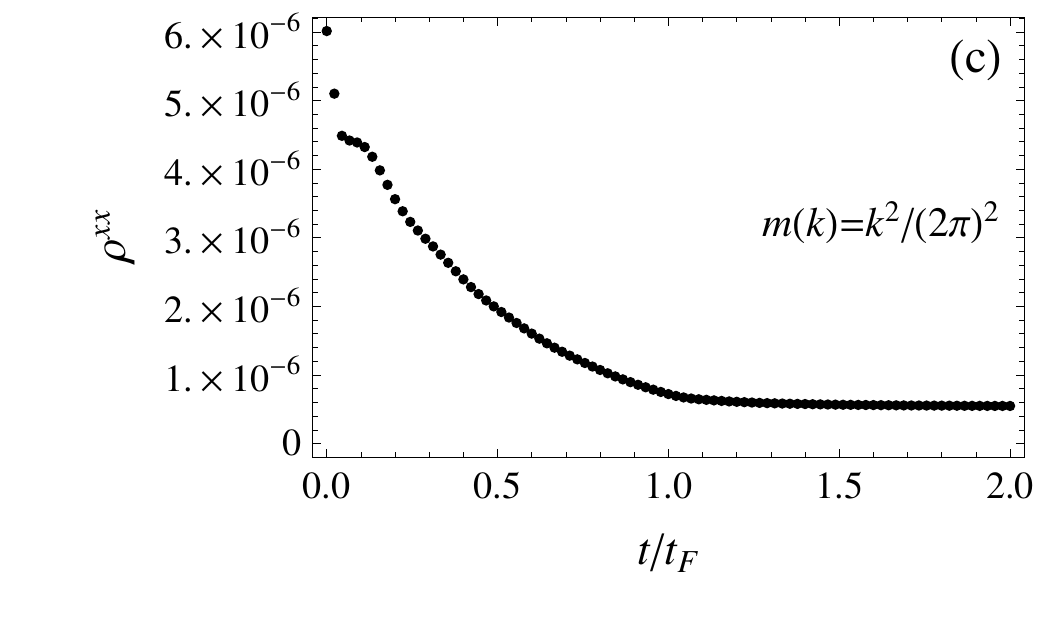}~\includegraphics[width=.49\textwidth]{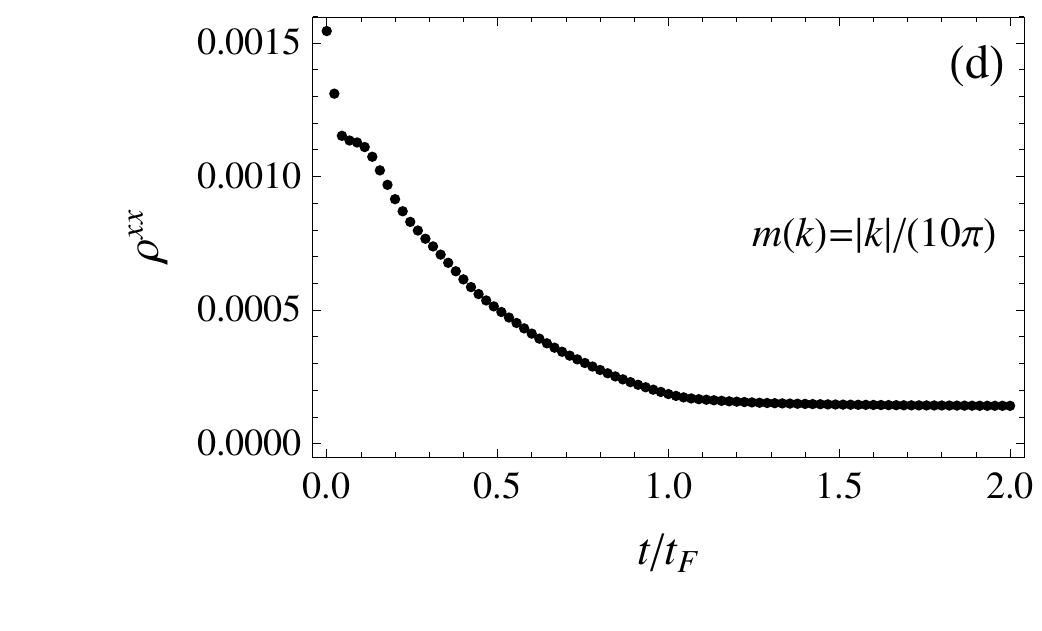}\\
\includegraphics[width=.49\textwidth]{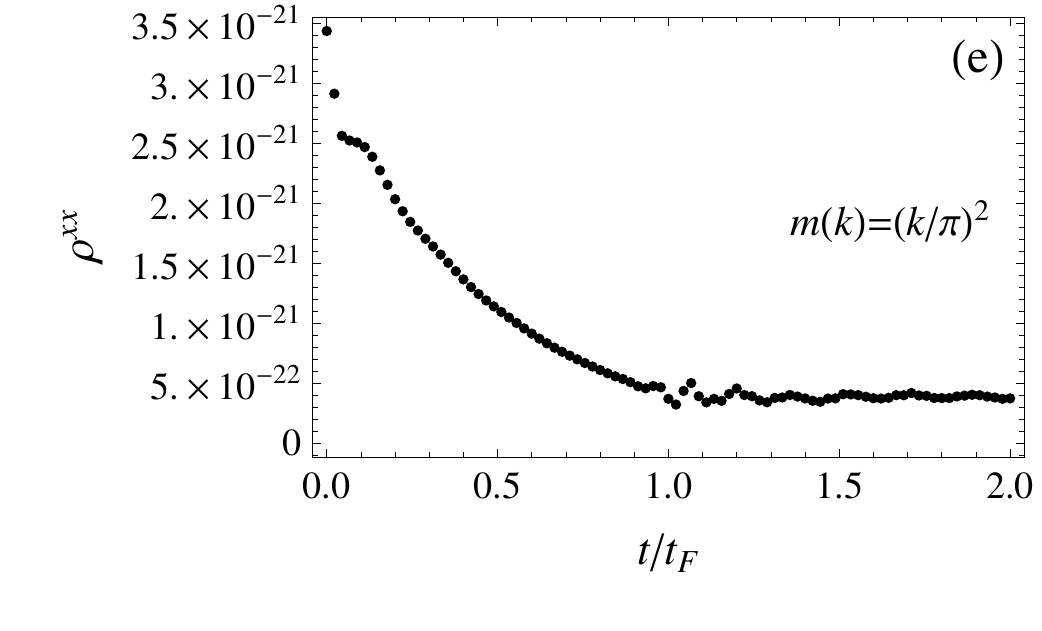}~\includegraphics[width=.49\textwidth]{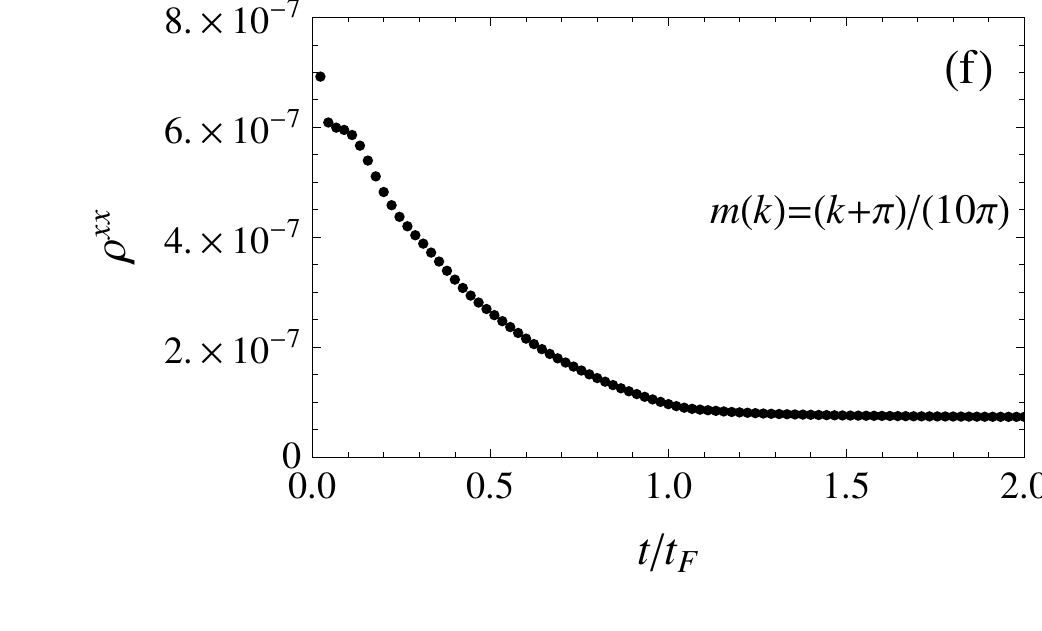}
\caption{
\small  Two-point longitudinal correlation function for $\ell=60$ as a function of time.
All figures refer to quenches from $h_0=1/3$ to $h=2/3$. 
Each panel corresponds to a different excited state with characteristic function $m(k)$ given by:
a)  $m(k)=\theta(\pi/2-|k|)$, b)  $m(k)=|k|/\pi$  c) $m(k)={k}^2/({2\pi})^2$,   d)  $m(k)=|k|/(10\pi)$, 
e) $m(k)=(k/\pi)^2$, f) $m(k)=(k+\pi)/(10\pi)$. 
}
\label{fig:graficiCx60}
\end{center} 
\end{figure} 

\begin{figure}[t] 
\begin{center} 
\includegraphics[width=.5\textwidth]{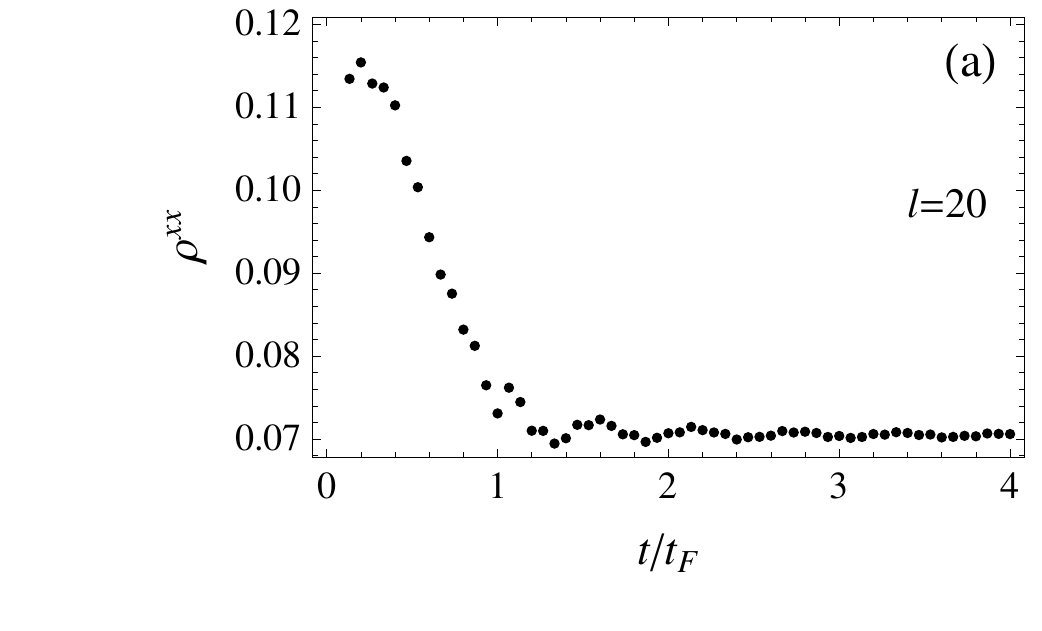}~\includegraphics[width=.49\textwidth]{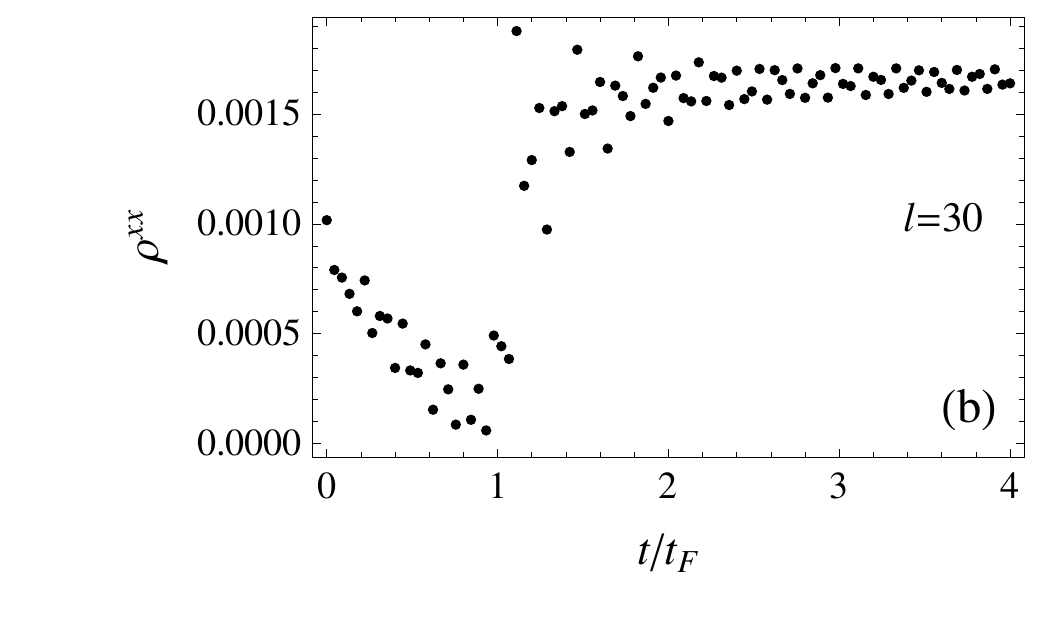}\\
\includegraphics[width=.5\textwidth]{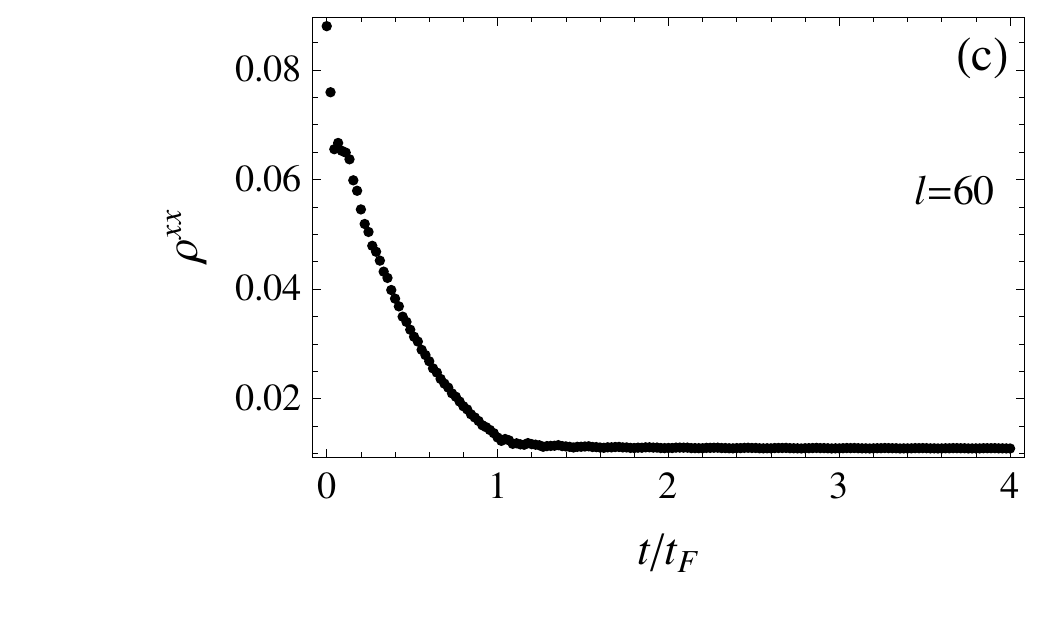}~\includegraphics[width=.49\textwidth]{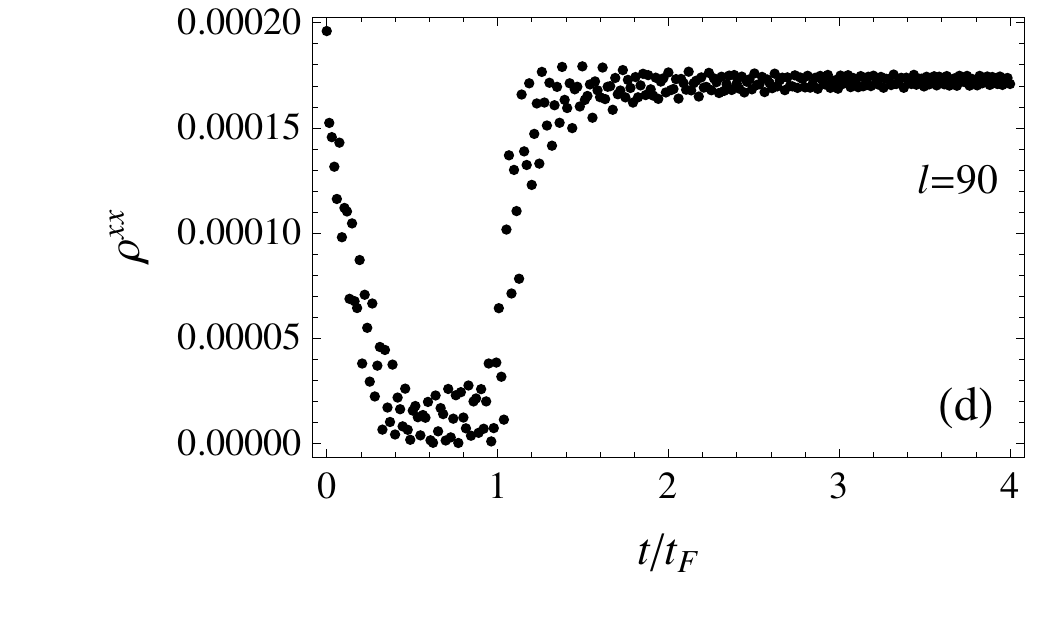}
\caption{\small  Two-point longitudinal correlation function for the quench from $h_0=1/3$ to $h=2/3$ for the 
initial excited state with $m(k)=\theta(k-\pi/2)$. 
The four panels correspond to different distances, namely: a) $\ell=20$,  b) $\ell=30$,   c) $\ell=60$, d) $\ell=90$.
}
\label{fig:graficiCxteta}
\end{center} 
\end{figure}

We report in this section the numerical results obtained for the longitudinal correlator for various initial excited states. 
We restrict ourselves to quenches within the ferromagnetic phase, because, as for the ground-state case \cite{cef,cef-i}, 
quenches between the phases and within the paramagnetic phase have a more complicated time dependence. 
All the following numerical results have been obtained for the quench from 
$h_0={1}/{3}$ to $h={2}/{3}$, but the conclusions we draw are valid for arbitrary quenches 
within the ferromagnetic phase. 
The time evolution of the two-point function $\rho^{xx}(\ell,t)$ is reported in units of the Fermi time $t_F$, defined as \cite{cef-i}
\begin{equation}
t_F=\frac{\ell}{2 v_{\rm max}},
\end{equation} 
where $v_{\rm max}$ is the maximal propagation velocity of the elementary excitations
\begin{equation}
v_{\rm max}=\max_{k\in [-\pi,\pi]}|\epsilon'_k|=\mathrm{min}[h,1].
\end{equation}
In Fig. \ref{fig:graficiCx60} we report the obtained numerical results for the correlation function at fixed distance 
$\ell=60$. All the data reported in these plots show a quite general behaviour:
for $t<t_F$ the correlation function decays exponentially, while for $t>t_F$ it shows a slow relaxation 
toward the GGE value. 
This is a manifestation of the light-cone spreading of correlations \cite{cc-06} also for these quenches from excited states. 
However, not all the initial excited states we analysed behave in this way and 
for that reason we show it separately in Fig. \ref{fig:graficiCxteta}.
There we report the time evolution from the state characterised by $m(k)=\theta(k-\pi/2)$  which appears qualitatively 
different from the others: while for $\ell=20, 60$ it is similar to the other cases in Fig. \ref{fig:graficiCx60}, 
for $\ell=30,90$ after an initial decay the correlation function displays a sort of plateaux and at $t\sim t_F$ 
it sets around the GGE value which is reached in an oscillating manner. 
It is not clear to us what the physical phenomenon behind this behaviour is, 
but we have observed it only for NPIS (i.e. for $m_k\neq m_{-k}$).
Furthermore, the analytic result obtained in the next subsection for PIS shows that they always behave as in Fig. \ref{fig:graficiCx60}.
Thus it is natural to believe  that the anomalous behaviour in Fig.  \ref{fig:graficiCxteta} is due to the non-vanishing odd 
conservation laws during the time evolution, but how exactly this happens is still to be understood, 
even because some NPIS behaves like their PIS counterpart as the case (f) in Fig. \ref{fig:graficiCx60}.

Also for the well behaved cases  in Fig. \ref{fig:graficiCx60} there is a fundamental difference 
compared to the ground-state initial case which is worth mentioning. 
In equilibrium, all these states are characterised by a vanishing one-point function $\braket{\sigma^x}$
explaining why the initial values of the  two-point functions in  Fig. \ref{fig:graficiCx60}
is always within the range $10^{-1}-10^{-27}$ (and obviously it goes to zero increasing $\ell$), 
while it was close to $1$ for the ground-state quench. 
Furthermore it also seems that the more excited the state is, the lower is the initial value of two point correlator.
This can be seen from Fig. \ref{fig:graficiCx60} by comparing the initial states in which $m_k$ has the same analytical form but different 
pre-factor.
However, we did not explore this aspect in detail because it is not of direct interest to this manuscript.

\subsection{Analytical full time evolution  for parity invariant states}

In this section we provide an analytic result for the time dependence of the longitudinal two-point function for PIS  
and quenches within the ferromagnetic phase. 
For a quench starting from the ground-state and within the ferromagnetic phase, 
the two-point correlation function in the space-time scaling limit is  \cite{cef,cef-i}  
\begin{align}
\label{formula}
 \rho^{xx}(\ell,t)\simeq & \;C^x \exp  \left[ \ell \int_0^\pi \frac{dk}{\pi}\ln [|\cos \Delta_k|]\theta(2|\epsilon'_k|t-\ell)
 +2t\int_0^\pi\frac{dk}{\pi}|\epsilon'_k|\ln[|\cos \Delta_k|]\theta(\ell-2|\epsilon'_k|t)\right],
\end{align}
where $\theta(x)$ is the Heaviside step function and $C^x$ is a coefficient which can also be calculated \cite{cef-i,cef-ii}.
This result is based on the multi-dimensional stationary phase approach developed in Refs. \cite{fc-08,cef,cef-i} and 
reported for completeness in Appendix \ref{AppA}.

The derivation of Eq. (\ref{formula}) is based on the fact that the $2\times2$ symbol $\hat{\Gamma}(k)$ can be cast into the form
(\ref{eq:symbol}) of Appendix \ref{AppA} which, in particular, implies that the block symbol is traceless. 
The symbol for the excited state $\hat{\Gamma}^e(k)$ given in Eq. \eqref{Gamma_k} is characterised by
\begin{equation}
\mathrm{Tr}[\hat{\Gamma}^e(k)]=2i m^A_k,\quad\quad\quad 
\det[\hat{\Gamma}^e(k)]=(m^S_k)^2-(m^A_k)^2=(1-2m_k)(1-2m_{-k}).
\end{equation}
Hence a generalised version of (\ref{formula}) can be derived only provided that the symbol is traceless, i.e. for PIS. 
In this case the symbol for the excited state is proportional to the one for the ground state $\hat{\Gamma}^{\text{gs}}(k)$,
indeed from Eq. (\ref{Gamma_k}) we have 
\begin{equation}
\hat{\Gamma}^e_{\rm PIS}(k)=-m_k^S\hat{\Gamma}^{\text{gs}}(k),
\end{equation}
and hence the coefficients $n_x$ and $\vec{n}_\bot$ appearing in Eq. (\ref{eq:symbol}) are 
\begin{equation}\label{symbol_even}
n_x=-m_k^S \cos{\Delta_k}, \quad\quad |\vec{n}_\bot|^2=\sin{\Delta_k}^2(m_k^S)^2.
\end{equation}
\begin{figure}[t!] 
\begin{center} 
\includegraphics[width=.49\textwidth]{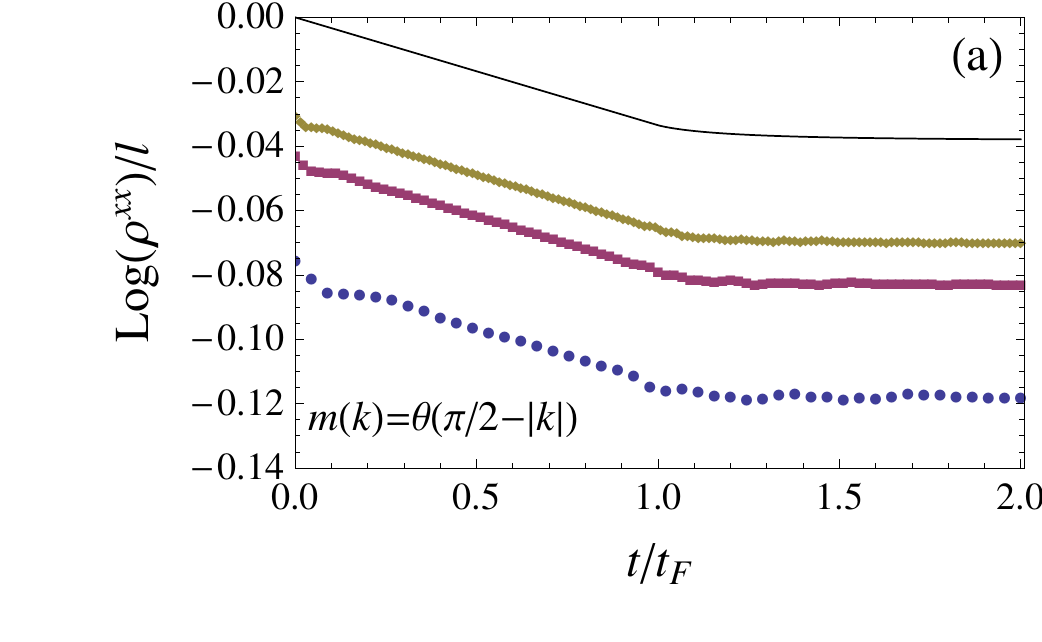}~\includegraphics[width=.49\textwidth]{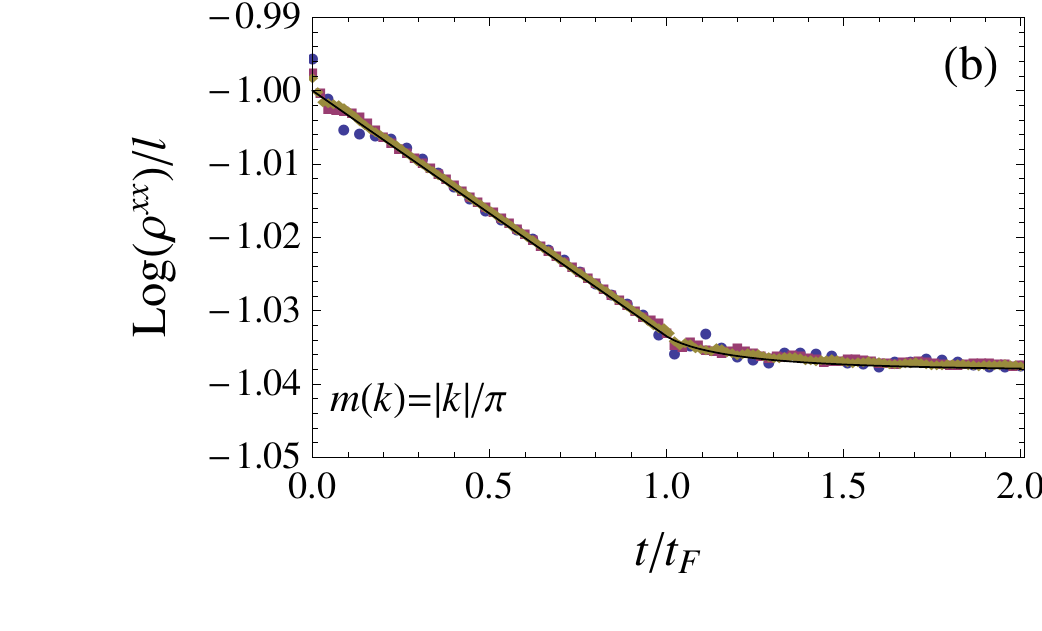}\\
\includegraphics[width=.49\textwidth]{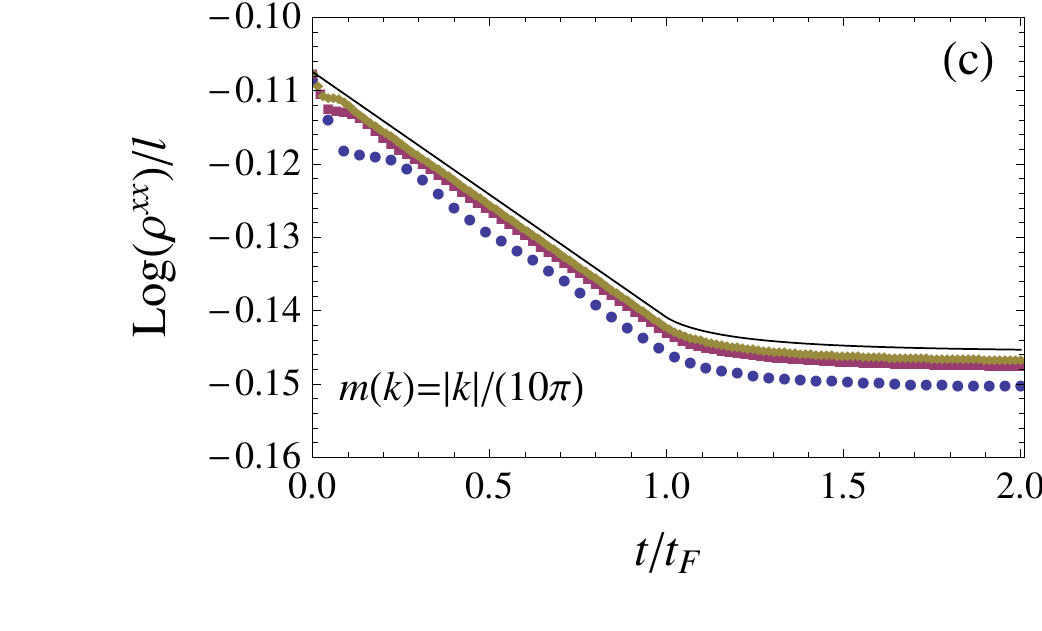}~\includegraphics[width=.49\textwidth]{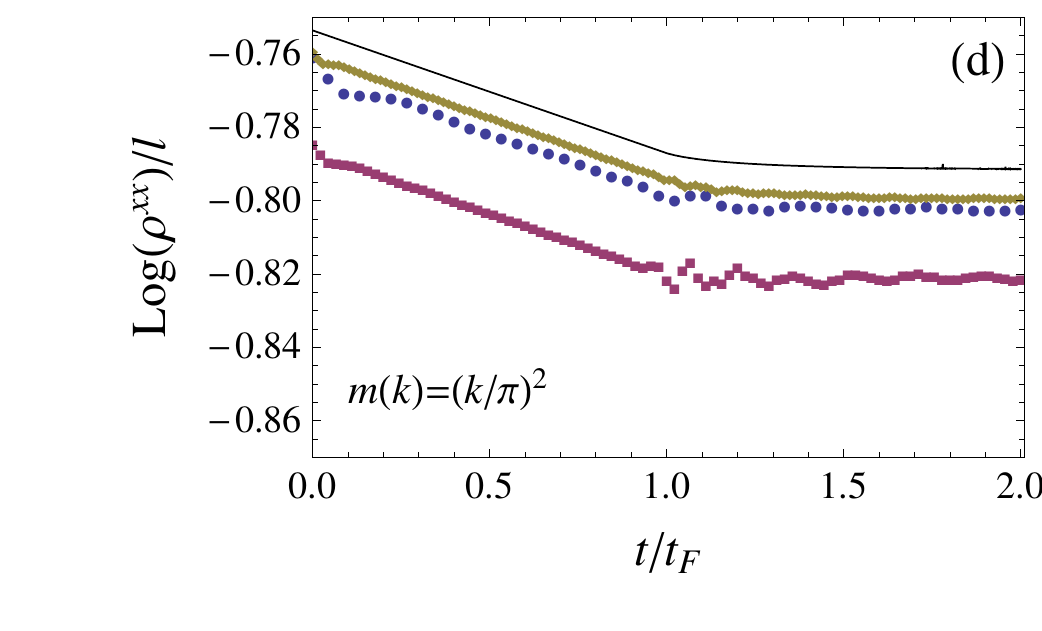}\\
\caption{\small Scaling behaviour of $\ln (\rho^{xx})/\ell$ vs $t/t_F$. 
The continuous line is the analytical prediction in Eq. (\ref{formula2}), 
the blue points  correspond to $\ell=30$, the violet squares to $\ell=60$ and the dark green diamonds to $\ell=90$.
The various panels corresponds to different initial states with: 
a) $m(k)=\theta(\pi/2-|k|)$, b)  $m(k)=|k|/\pi$,  c) $m(k)=|k|/(10\pi)$, d) $m(k)=(k/\pi)^2$.
} 
\label{fig:graficiCxana}
\end{center} 
\end{figure} 

At this point, the generalisation of Eq. (\ref{formula}) to excited initial states is a straightforward application of 
Eq. (\ref{maufor}) in the appendix which leads to
\begin{align}
\label{formula2}
 \rho^{xx}_{m_k}(\ell,t)\simeq \; C_{m_k}  &
 \exp \left[ \ell \int_{-\pi}^\pi \frac{dk}{2\pi} \left(1-2|\epsilon'_k|\frac{t}{\ell}\right)\ln(|m_k^S|)
 \theta(\ell-2|\epsilon'_k|t)\right]\nonumber\\
 &\exp \left[ \ell \int_{-\pi}^\pi \frac{dk}{2\pi}\ln[| \cos \Delta_k m_k^S|]\theta(2|\epsilon'_k|t-\ell)\right]\nonumber\\
  & \exp\left[2t\int_{-\pi}^\pi\frac{dk}{2\pi}|\epsilon'_k|\ln [ |\cos \Delta_k m_k^S|]\theta(\ell-2|\epsilon'_k|t)\right].
\end{align}
Notice that compared with the ground state result there is an important qualitative difference given by
the first line of the expression above that is absent only if $m_k=0$ identically, i.e. for the ground state. 
In the multidimensional stationary phase approach this term arises from to the fact that $n_x^2+|\vec{n}_\bot^2|\neq 1$.
This term is also responsible for an exponential decay in the distance $\ell$ of the correlation function 
in the initial state, a fact that we anticipated in the previous section and that the above result proves.

In Fig. \ref{fig:graficiCxana} we report the numerically calculated correlation functions and we compare with the 
analytic prediction  (\ref{formula2}). 
We plot the logarithm of the correlation in order to see clearly the exponential decay for $t<t_F$ followed by 
a slow relaxation for $t>t_F$. 
It is evident that increasing $\ell$, the various curves approach the asymptotic result in Eq. (\ref{formula2}). 
Finite size (in $\ell$) effects are almost exclusively due to the undetermined constant $C_{m_k}$
which in this kind of plots produces a $\ell^{-1}$ time-independent finite size correction, as
proven by the fact that all curves are basically parallel.  
The behaviour of the state characterised by $m_k=(k/\pi)^2$ is a bit peculiar because increasing $\ell$ the numerical curves 
approach the analytic result in a non-monotonic way (the result for $\ell=30$ is in between those for  $\ell=60$ and $\ell=90$). 
This is not at all surprising because the coefficient $C_{m_k}$ can depend on $\ell$ in an oscillating way 
every time that the symbol is a non-analytic function (for example in the long time limit the strong Szeg\H{o}'s lemma 
needs to be generalised to the Fisher--Hartwig formula, see e.g. \cite{cef-ii} for explicit examples).

While in principle it is possible to compute the coefficient $C_{m_k}$ for every excited states, 
each of them requires a different calculations and it is not worth analysing all of them. 
In order to give a typical example, in the next subsection we calculate this pre-factor for the state  $m_k=k^2/(2\pi)^2$.

\subsubsection{Computation of the pre-factor for an initial excited state}

In this subsection we  compute the coefficient $C_{m_k}$ for the state  $m_k=k^2/(2\pi)^2$. 
In general, the coefficient $C_{m_k}$ can be extracted by evaluating it at infinite time when the matrix $\Gamma$
becomes a standard  time independent Toeplitz one and we can apply Szeg\H{o}'s lemma  or generalisations 
(assuming that the space-time scaling limit and direct $t\to\infty$ commute, as it can be checked a posteriori). 
Calculations are largely simplified when the symbol is a smooth function and the strong Szeg\H{o}'s lemma holds. 
For the cases explicitly reported in the previous subsection, this happens only for $m_k={k}^2/({2\pi})^2$ 
and we will see that a closed form for  $C_{m_k}$ can indeed be found easily. 
On the contrary, for the other states examined above the symbol is not smooth and, as a consequence,  
generalisations of the Szeg\H{o}'s lemma are necessary, but since they require a case by case examination we 
prefer not to go into such details.

\begin{figure}[t] 
\begin{center} 
\includegraphics[width=.49\textwidth]{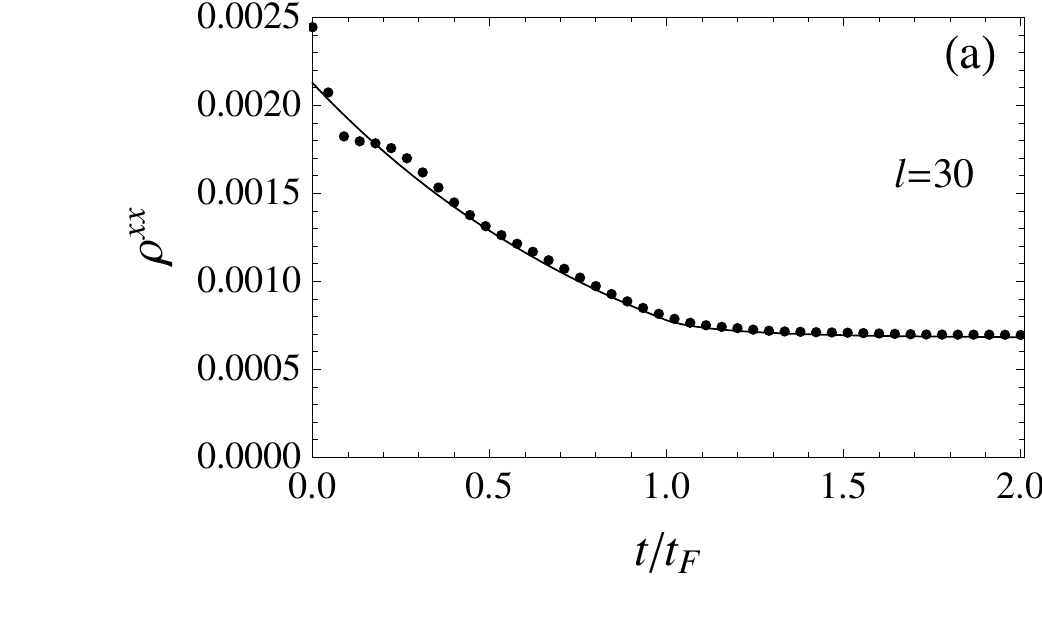}~\includegraphics[width=.49\textwidth]{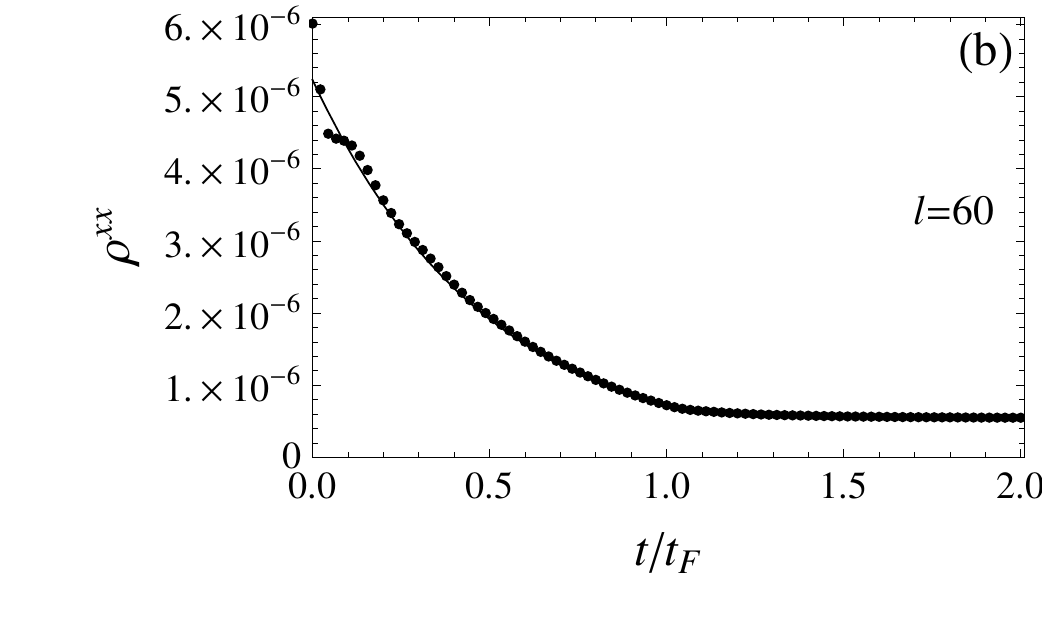}
\caption{\small Time-dependent correlation function $\rho^{xx}(\ell,t)$ for the quench from $h_0=1/3$ to $h=2/3$ 
with initial state given by $m(k)={k}^2/({2\pi})^2$.
The points are the numerical determination of the Pfaffian and the continuous line is the analytic prediction in Eq. (\ref{formula2})
with pre-factor $C_{m_k}$ fixed by Eq.  (\ref{Cpar}).
The two panels correspond to $\ell=30$ (a) and  $\ell=60$ (b).
The agreement is excellent in both cases.} 
\label{fig:graficiCx3060ana}
\end{center} 
\end{figure}

As shown in Refs. \cite{SZ,cef-ii}, the strong Szeg\H{o}'s lemma gives the pre-factor $C_{m_k}$ in the form
\begin{equation}\label{defCX}
C_{m_k}=\exp{\left[\sum_{q\geq 1}{q ({\ln t}_{\infty})_q ({\ln t}_{\infty})_{-q}}\right]},
\end{equation}
where $({\ln t}_{\infty})_q$ is the $q$-th coefficient of the Fourier expansion of the symbol at infinite time, i.e.
\begin{equation}
({\ln t}_{\infty})_q=\int_{-\pi}^{\pi}\frac{dk}{2\pi}({\ln t}_{\infty}(e^{ik})) e^{-ikq}.
\end{equation}
For  $m(k)={k}^2/({2\pi})^2$ we have
\begin{equation}
t_\infty (e^{ik})=\cos\Delta_k \left(1-\frac{k^2}{2 \pi ^2} \right),
\end{equation}
and hence the Fourier coefficients are
\begin{equation}\label{Fcoeff}
(\ln t_\infty)_q=\left\{\begin{array}{ll} \frac{h_0^q-2h_1^{-q}}{2q}-\frac{2 i\pi \sin(\pi q)}{\sqrt{2\pi}q}, &\quad \mathrm{if}\quad q>0,\\
\frac{2 h_1^q-2h^{-q}-h_0^{-q}}{2q}-\frac{2 i\pi \sin(\pi q)}{\sqrt{2\pi}q}, &\quad \mathrm{if}\quad q<0,
\end{array}\right.
\end{equation}
where 
\begin{equation}
h_1=\frac{1+h h_0+\sqrt{(h^2-1)(h_0^2-1)}}{h+h_0}.
\end{equation}
Computing the sum in Eq. (\ref{defCX}), all the pieces depending on $\sin(\pi q)$ in (\ref{Fcoeff}) cancel, leading to
\begin{equation}
C_{m_k}=\frac{(h-h_1)(h_0-h_1)}{\sqrt{1-h h_0}(1-h_0^2)^{1/4}(h_1^2-1)}.
\label{Cpar}
\end{equation}
Remarkably, this is independent on the specific value of $m(k)$ and it is indeed the same value obtained for 
the initial ground state \cite{cef-ii}. 
In Fig. \ref{fig:graficiCx3060ana} we compare with the numerical results the full prediction for the time-dependent correlation function in Eq. (\ref{formula2})
with the pre-factor given by Eq (\ref{Cpar}), hence with no unknown  parameter. 
The agreement between the analytic formula  and the numerics  is excellent.

\section{Time evolution of the entanglement entropy}
\label{enta}

In this section we turn to the study of the time evolution of the entanglement entropy of a block of $\ell$ contiguous spins after the 
quench from an excited state of the pre-quench Hamiltonian.

As explained in Sec. \ref{sec.org}, the time evolution of the entanglement entropy can be obtained from the 
eigenvalues of the correlation matrix $\Pi$ in Eq. (\ref{Ppi}).
For the case we are interested in the constituent blocks of this matrix have the form
\be 
\Pi_n=\left(
\begin{array}{cc}
h_n & g'_n\\
-g'_{-n} & f_n
\end{array}
\right),
\end{equation}
where  $f_n$, $h_n$ are the ones respectively given in Eq. (\ref{fdef}) and Eq. (\ref{hdef}), while $g'_n$ turns out to be
\begin{equation}
g'_n=\int_{-\pi}^{\pi}\frac{dk}{2\pi}e^{ikn}\left[- m^S_k e^{i \theta(k)}(\cos{\Delta_k}-i\cos{2t\epsilon_k}\sin{\Delta_k})\right],
\end{equation}
hence it differs from (\ref{gdef}) by a factor $e^{ik}$.

By numerically calculating the eigenvalues of the matrix $\Pi$ and inserting them in Eq. (\ref{SvsH})
 we obtain a numerical estimate of the entanglement entropy.

\subsection{Analytic evaluation of the entanglement entropy for parity invariant initial states}

In this subsection we generalise the analytical formula  which describes the time-dependence of the entanglement entropy 
of a block of spins of length $\ell$ after a quench starting from the ground state 
\cite{cc-05,fc-08}, in the thermodynamic limit and in the limit of a large block $\ell\gg 1$. 
The leading behaviour in the space-time scaling limit for a quench from the ground state is \cite{fc-08}
\begin{equation}
S_A(t)=2t\int_{2|\epsilon'_k |t<\ell}\frac{d k}{2\pi}|\epsilon'_k | H(\cos{\Delta_k}) +
\ell\int_{2|\epsilon'_k |t>\ell}\frac{d k}{2\pi} H(\cos{\Delta_k}),
\label{Sground}
\end{equation}

In the case of a traceless $2\times2$ symbol, i.e. for PIS, the generalisation of the aforementioned formula is direct
using Eq. (\ref{maufor}) in Appendix \ref{AppA}. 
However, for NPIS a closed form cannot be obtained because the proof of the previous formula crucially relies on 
the tracelessness of the symbol.

Eq. (\ref{maufor}) can be applied to the entanglement entropy since Eq. (\ref{SvsH}) is equivalent to  
\begin{equation}
S=\mathrm{Tr}[H[\Pi]],
\end{equation}
where $H(x)$ is given in Eq. (\ref{SvsH}).
Using Eq. (\ref{maufor}) we have
\begin{align}
\lim_{t,\ell \to \infty \atop t/\ell= {\rm const}}\frac{\mathrm{Tr} [H[\Pi]]}{\ell}&= 
\int_{-\pi}^ \pi \frac{d k}{2\pi}\mathrm{max}\left(1-2|\epsilon'_k|\frac{t}{\ell},0\right) H\left(\sqrt{n_x (k)^2+|n_\bot (k)|^2}\right)+\nonumber\\
&+\int_{-\pi}^ \pi\frac{d k}{2\pi}\mathrm{min}\left(2|\epsilon'_k|\frac{t}{\ell},1\right) H\left(n_x(k)\right).
\end{align}
Inserting in this equation the explicit expressions for $n_x(k)$ and $n_\perp(k)$  in Eq.  
(\ref{symbol_even}) we get, in the scaling limit, the entanglement entropy 
\begin{align}
\label{Sperspin}
S_A(\ell,t) \simeq
& \int_{-\pi}^\pi \frac{dk}{2\pi} \left(\ell-2|\epsilon'_k| t\right) H(m_k^S) \theta(\ell-2|\epsilon'_k|t)+\nonumber\\
 &+ \ell\int_{-\pi}^\pi \frac{dk}{2\pi}H\left[m_k^S\cos \Delta_k\right]\theta(2|\epsilon'_k|t-\ell)+\nonumber\\
  &+ 2t \int_{-\pi}^\pi\frac{dk}{2\pi}|\epsilon'_k| H\left[m_k^S\cos \Delta_k\right]\theta (\ell-2|\epsilon'_k|t).
\end{align}
Also the entanglement entropy shows a light-cone behaviour, i.e. a linear growth for $t<t_F$
followed by a slow saturation. However, 
even in this case there is a main qualitative difference with the ground state result in Eq. (\ref{Sground})
which is represented by the first line of the equation. 
This is again technically due to the fact that $n_x^2 +n_\perp^2\neq 1$ and physically reflects the property that the entanglement entropy in the initial state is extensive.
We mention that the zero time limit agrees
with the results for the entanglement entropy found in Ref. \cite{afc-09} for the same class of excited states, 
but with a different method. Thus, for the entanglement entropy, the limit $t\to0$ and the space-time scaling 
limit turn out to commute.

\begin{figure}[t!] 
\begin{center} 
\includegraphics[width=.49\textwidth]{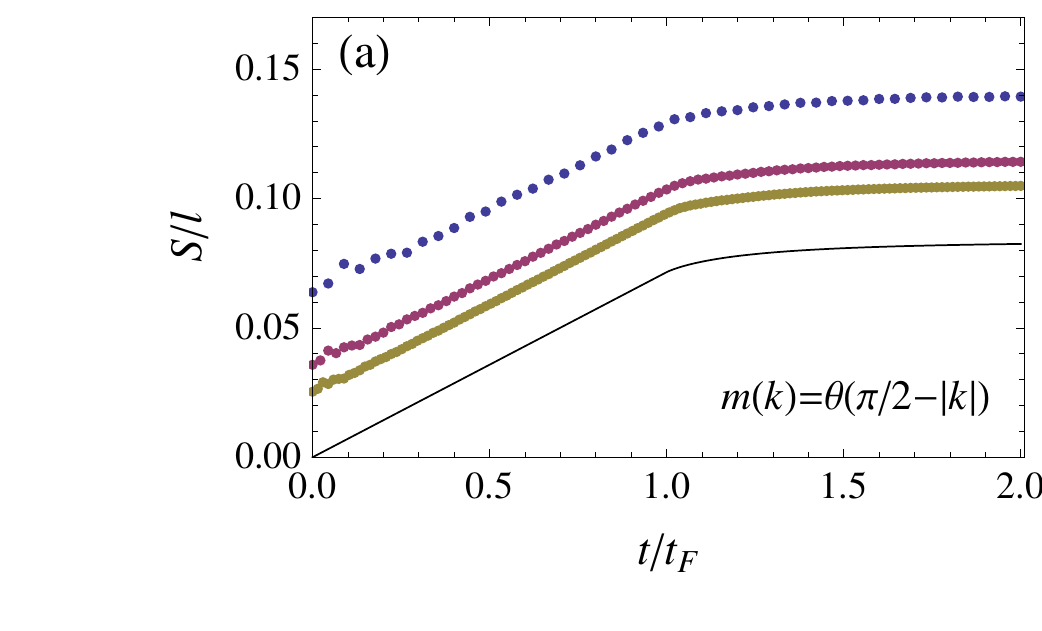}~\includegraphics[width=.49\textwidth]{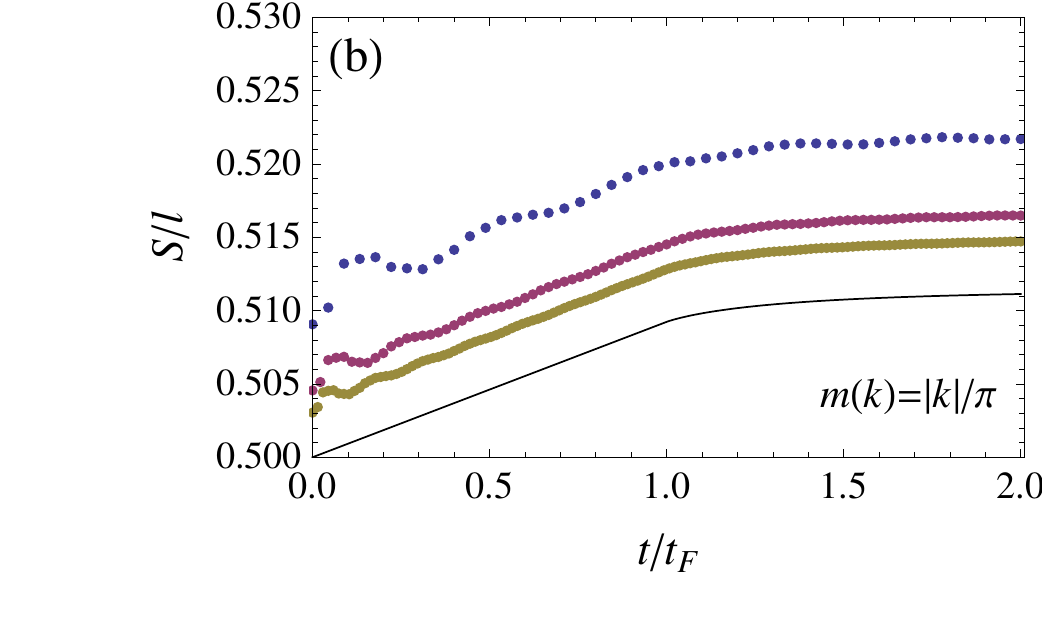}\\
\includegraphics[width=.49\textwidth]{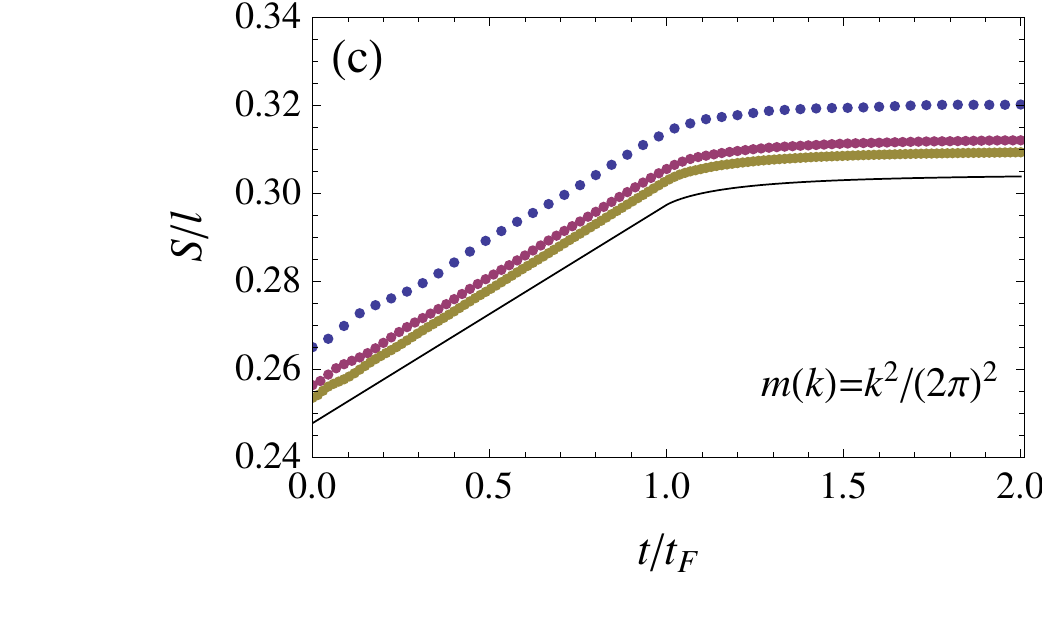}~\includegraphics[width=.49\textwidth]{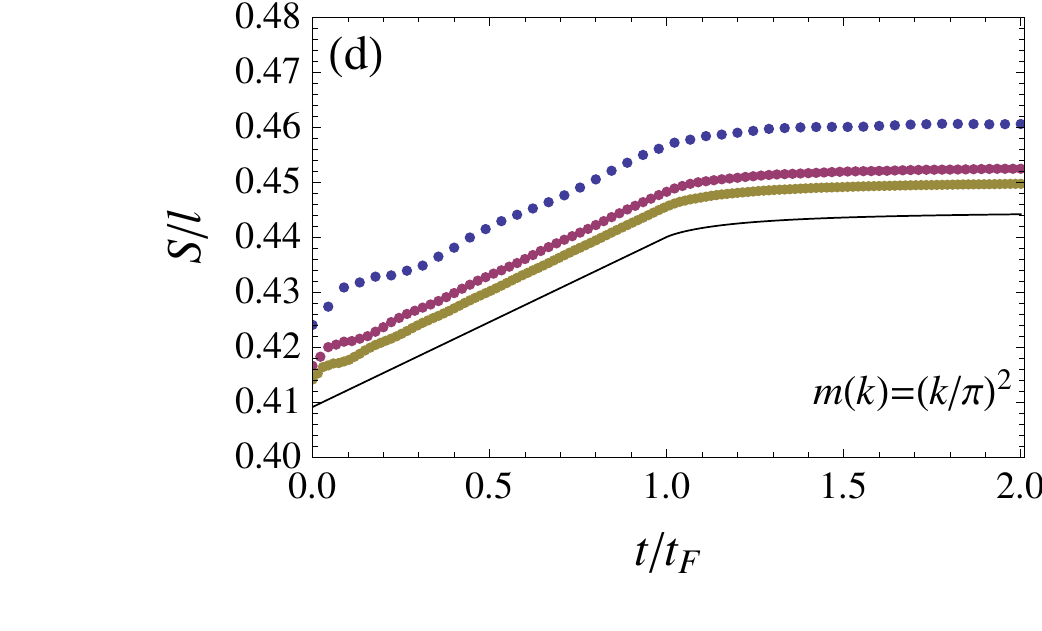}
\caption{\label{fig:entangl}\small Time dependence of the entanglement entropy per spin starting from parity invariant states. 
The continuous line is the analytical formula Eq. \eqref{Sperspin}, the blue points correspond to $\ell=30$, the violet squares to $\ell=60$ 
and the dark green diamonds to $\ell=90$.
(a) $m(k)=\theta(\pi/2-|k|)$, (b)  $m(k)=|k|/\pi$, (c) $m(k)=k^2/(2\pi)^2$, (d) $m(k)=(k/\pi)^2$. 
}
\end{center} 
\end{figure}

\begin{figure}[t] 
\begin{center} 
\includegraphics[width=.49\textwidth]{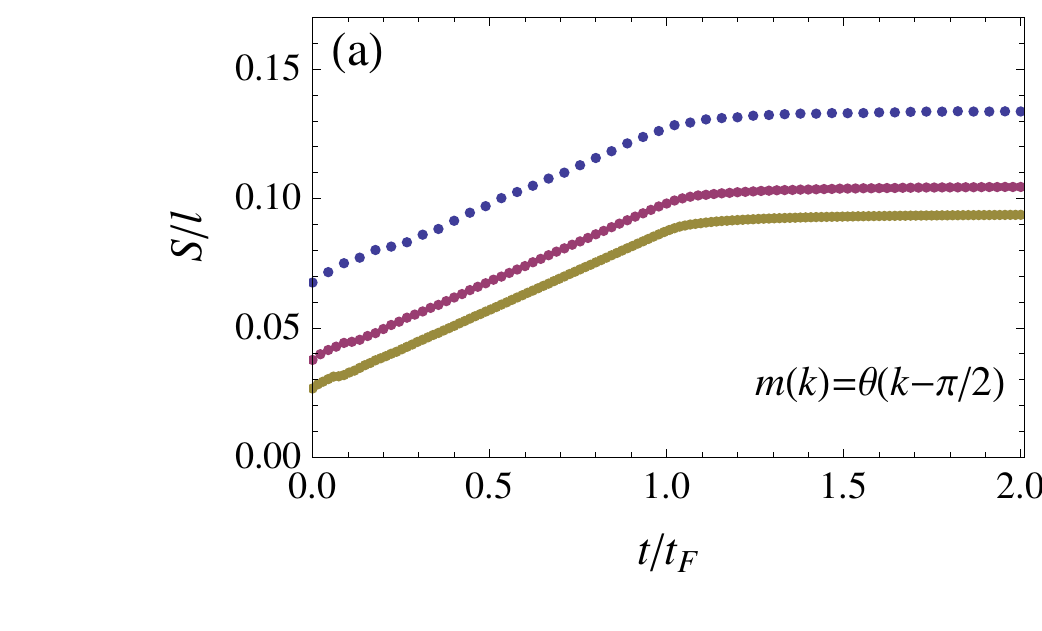}~\includegraphics[width=.49\textwidth]{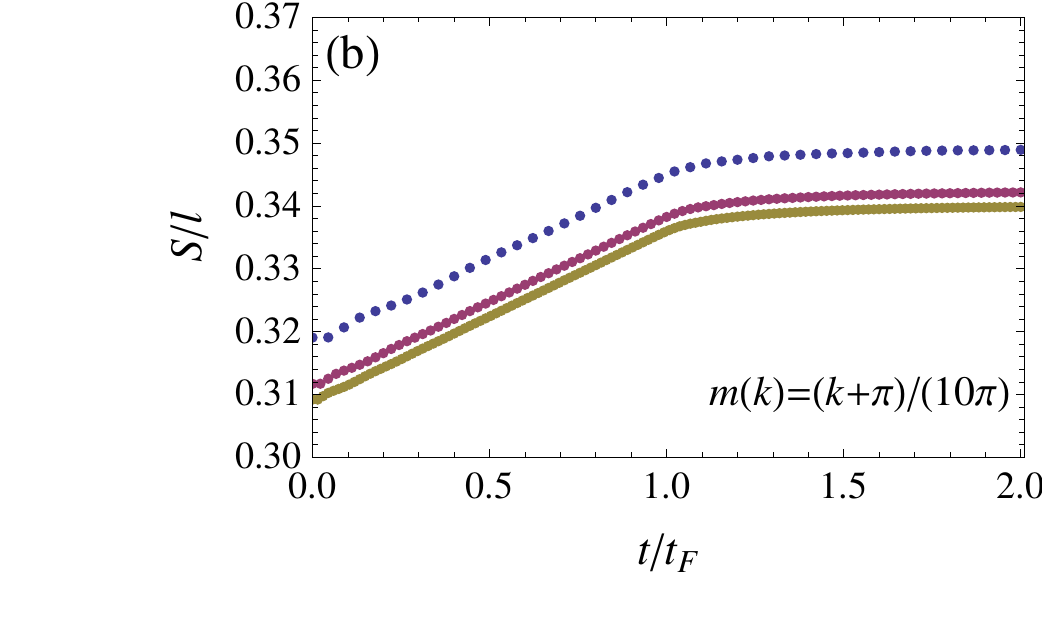}
\caption{\label{fig:entanglb}\small Time dependence of the entanglement entropy per spin for two non-parity invariant states. 
The blue dots correspond to $\ell=30$, the violet ones to $\ell=60$ and the dark green ones to $\ell=90$.
(a): $m(k)=\theta(k-\pi/2)$.  (b): $m(k)=(k+\pi)/(10\pi).$
 } 
\end{center} 
\end{figure}

In order to show the correctness of our prediction, we report in Fig. \ref{fig:entangl} the numerical results for the entanglement 
entropy per spin starting from a few different parity invariant initial states which are compared with the analytic prediction (\ref{Sperspin}). 
Increasing the size of the block of spins, the entanglement entropy obtained with the determinant approach gets closer 
and closer to the analytic formula. Being the various finite $\ell$ results all parallel to the prediction it is clear 
that the leading finite-size correction is just an additive constant which in principle could be 
obtained by means of Szeg\H{o}'s lemma or generalisations thereof.

In Fig. \ref{fig:entanglb} we report the time evolution of the entanglement entropy per spin starting from non-parity invariant states. 
The light-cone spreading  of the correlation is clear also in this case, but we do not have an analytic prediction. 
In order to exclude simple generalisations of Eq. (\ref{Sperspin}), we also checked that the prediction 
for parity invariant states (using only the $m^S_k$ part of the state) does not describe the numerical results.

\subsection{Infinite time limit of the entanglement entropy}

It is relatively easy to obtain the infinite time limit, not only for parity invariant states, but 
 for an arbitrary initial state. 
Indeed, for infinite time, the entanglement entropy can be written as \cite{jk-04,cc-05}
\begin{equation}
S_A=\frac{1}{4\pi i}\oint d \lambda H(\lambda)\frac{d}{d \lambda}\ln D_\ell(\lambda),\quad\quad 
D_\ell(\lambda)=\mathrm{det} (i\lambda I_\ell-\Pi_\ell),
\label{Sdet}
\end{equation}
where $I_\ell$ is the $2\ell \times 2\ell$ identity, $H(x)$ is defined in Eq. (\ref{SvsH}), and the integral is evaluated over 
a contour that encircles the segment $[-1,1]$. 
The asymptotic (in $\ell$) behaviour of  $\ln D_\ell$ is found using a generalisation of Szeg\H{o}'s lemma \cite{am-74}
 \begin{equation}
 \ln D_\ell(\lambda)=\frac{\ell}{2\pi}\int _{-\pi}^{\pi}d k \ln\left[\mathrm{det}\tilde{\Pi}(k)\right]+O(\ln \ell),
\label{Dlam}
 \end{equation}
where 
\begin{equation}
\tilde{\Pi}(k)=i\lambda I_1-\Pi(k)=\left(
\begin{array}{cc}
i\lambda -im_k^A & m_k^S e^{-i\theta(k)}\cos\Delta_k\\
-m_k^S e^{i\theta(k)}\cos\Delta_k  & i\lambda -i m_k^A
\end{array}
\right),
\end{equation}
where $\Pi(k)$ is given by the infinite time limit  of Eq. (\ref{Pitop}). 
Inserting the expression for $\det \tilde{\Pi}(k)$ in Eq. (\ref{Dlam}) and then in Eq. (\ref{Sdet}) we have the  
linear part in $\ell$ of the entanglement entropy 
\bea
S_A(\ell,t=\infty)&\simeq&\frac{\ell}{2\pi} \int_{-\pi}^{\pi} dk
\frac{1}{4\pi i}\oint d \lambda H(\lambda) 
\frac{2(\lambda-m_k^A)}{(\lambda-m_k^A)^2-(m_k^S)^2\cos ^2 \Delta_k}
\nonumber\\ &=& \ell \int_{-\pi}^{\pi} \frac{dk}{2\pi} H(m^A_k +m^S_k \cos\Delta_k),
\eea
where in the last line we first shifted the integral by $m^A_k$ and then used the residue theorem.
This is allowed because for arbitrary $m_k$ we have $-1<m^A_k +m^S_k \cos\Delta_k<1$, 
where the function $H(x)$ is real.  
We stress that this is true for arbitrary states and not only for PIS. 
When specialised to PIS, the argument of the integral above reduces to $H(m^S_k \cos\Delta_k)$ 
which is the long time limit of Eq. (\ref{Sperspin}) showing that for the entanglement entropy the scaling limit and the long time limit commute for arbitrary quenches.

\section{Conclusions}

We have considered the time evolution after a quench of the transverse magnetic field in the Ising model 
starting form an arbitrary excited state of the pre-quench Hamiltonian having the form 
\be
|\Psi_0\rangle=|m_k\rangle=\prod_{k} ({b'}^\dag_k)^{m_k}|0\rangle. 
\label{psi02}
\ee 
This state is fully specified by the characteristic function $m_k=0,1$ in finite systems, which in the thermodynamic 
limit becomes an arbitrary function $m(k)$ with $k\in[-\pi,\pi]$ and  $m_k\in[0,1]$.  
It turned out that important quantitative and qualitative differences in the time evolution arise between 
parity invariant initial states (i.e. with $m_k=m_{-k}$ for all $k$) and non parity invariant ones. 

We showed that for an arbitrary state of the form (\ref{psi02}) the long time limit of any local observable 
can be evaluated by means of GGE. 
The proof is based on the equivalence of two-point fermion correlations in the GGE and in the long time limit 
(taken, as usual, after the thermodynamic limit). 
Wick's theorem then allows for the construction of the full reduced density matrix of any finite block of spins and hence any local multi-point correlation. 
Although we limited ourself to the study of equal-time quantities, the general result of Ref. \cite{eef-12} 
allows us to conclude that the GGE describes also different times correlations. 

Then we turned to the study of observables. We first considered the transverse magnetisation 
for which the non-parity invariance of the state does not play any role. 
We calculated the approach to the stationary value by means of the stationary phase approximation and 
we always found a power-law behaviour, but characterised by powers which depend on the initial state.

We then considered the two-point longitudinal correlation function at distance $\ell$, since the one-point function vanishes for states of the form (\ref{psi02}) even in the symmetry-broken phase. 
For parity invariant states and for quenches with the ferromagnetic phase, we found analytically 
the space-time scaling limit of this correlation by means of the multi-dimensional stationary phase approach.
In all cases, the correlation function displays a typical light-cone feature with 
exponential relaxation for $t<t_F= \ell/2v_{\rm max}$ and slow relaxation for $t>t_F$.
For non-parity invariant states, the numerical results show that the behaviour can be very different 
compared to the parity invariant counterpart and we still do not have a full understanding of the problem.

We also studied the entanglement entropy of a block of $\ell$ consecutive spins. 
We again found a light-cone spreading (i.e. linear increase followed by slow saturation)
for arbitrary quenches and for arbitrary initial states. 
Also in this case, the space-time scaling limit is obtained analytically by means 
of the multi-dimensional stationary phase approach, but again only for parity invariant states. 
The infinite time limit has been derived for an arbitrary initial state, independently of its parity. 

It is clearly an interesting open problem to understand the time-dependence of 
both the two-point correlation function and the entanglement entropy for non-parity invariant initial states. 
The different behaviour in these two classes of states could also be related to the fact that 
for non-parity invariant states the number of local conserved charges is doubled.

\section*{Acknowledgments}   
PC and MK acknowledge the ERC  for financial  support under Starting Grant 279391 EDEQS. MK acknowledges financial support from the Marie Curie IIF Grant PIIF-GA-2012-330076. 
We thank Maurizio Fagotti for very fruitful discussions.

\appendix

\section{The multidimensional stationary phase approximation}
\label{AppA} 

The evaluation of the correlation function $\rho^{xx}(\ell,t)$ and of the entanglement entropy $S_A(\ell,t)$ 
for large $\ell$ is equivalent to the asymptotic evaluation of the determinants and traces of a $2\times 2$ block Toeplitz matrices, 
i.e. like $\Pi$ and $\Gamma$ in  Eqs. (\ref{Ppi}) and (\ref{gamma}).
Several techniques like Szeg\H{o}'s lemma and the Fisher-Hartwig conjecture \cite{SZ} permit
the evaluation of traces/determinants of these matrices when the elements do not depend explicitly on the matrix size. 
This is in contrast to our case, where we are interested in the space-time scaling limit $\ell,t\to\infty$ with finite ratio $\ell/t$. 
Thus, each element of the matrices $\Gamma$ and $\Pi$ in the space-time scaling limit depends on a parameter (namely $t$)  
which is proportional to the matrix dimension $2\ell$. 
This precludes the application of the aforementioned techniques, except for $t=0$ and in the limit $t=\infty$. 
In order to deal with arbitrary large values of $t,$ in Refs. \cite{fc-08,cef,cef-i} a new approach based on a multi-dimensional 
stationary phase approximation was developed.
In this appendix we report  the main result of Ref. \cite{cef-i} which has been extensively used in the text.

In Ref. \cite{cef-i} a very general result was obtained for any 
$2\times 2$-block Toeplitz matrix $\Lambda$ with a symbol $\hat{t}(k)$ that can be cast in the form 
\be\label{eq:symbol}
 \tilde{t}(k)=n_x(k)\sigma_x^{(k)}+\vec{n}_{\bot}(k)\cdot\vec{\sigma}^{(k)} e^{2i \epsilon(k)t\sigma_x^{(k)}}, \quad\quad \vec{n}_{\bot}(k)\cdot \hat x=0.
\ee
Here the time $t$ is the only parameter proportional to the matrix size $2\ell$, $n_x,n_\perp$ are fixed but otherwise arbitrary and
$\sigma^{( k)}$ denotes a local rotation of the Pauli matrices 
\be\label{eq:rotationsigma}
\sigma_\alpha^{( k)}\sim i e^{i \vec w( k)\cdot \sigma}\sigma_\alpha e^{-i \vec w( k)\cdot \sigma}\, .
\ee
All block symbols in Eq.  (\ref{eq:symbol}) are traceless and have determinant equal to $n_x^2+|\vec{n}_\bot^2|$. 

Under the condition (\ref{eq:symbol}), the asymptotic value in the space-time scaling limit of the trace of an arbitrary 
analytic function $F(x)$ evaluated  on the matrix $\Lambda$ can be derived. 
The explicit result of Ref. \cite{cef-i} is 
\begin{align}
\lim_{t,\ell\to\infty\atop t/\ell\ {\rm const}}
\frac{ {\rm Tr} [F(\Lambda ^2)]}{2\ell}&=&
\int_{-\pi}^ \pi\frac{d k_0}{2\pi}\mathrm{max}\left(1-2|\epsilon'(k_0)|\frac{t}{\ell},0\right) F\left(n_x (k_0)^2+|n_\bot (k_0)|^2\right)+\nonumber\\
&&+\int_{-\pi}^ \pi\frac{d k_0}{2\pi}\mathrm{min}\left(2|\epsilon'(k_0)|\frac{t}{\ell},1\right) F\left(n_x(k_0)^2\right).
\label{maufor}
\end{align}

In this manuscript we applied this formula to the entanglement entropy and the longitudinal correlation.
Indeed they can be written as functions of the matrices $\Pi$ and $\Gamma,$ respectively as
\bea
S_A  &=&{\rm Tr} (H[\Pi])\,,\\
\ln (\rho^{xx})^2&= & \frac 12 {\rm Tr} (\ln \Gamma^2)\,.
\eea
The function $H(x)$ is an analytic even function of $x$ for $x\in(-1,1)$ where the eigenvalues of $\Pi$ lie, 
and so Eq. (\ref{maufor}) can be applied with the only limitation that the symbol satisfies the constrain (\ref{eq:symbol}).
The function $\ln(x^2)$ is instead non-analytic in $x=0$ and this gives problems when the eigenvalues of $\Gamma$
approach $0$ in the thermodynamic limit. 
As discussed in Ref. \cite{cef-i}, this problem limits the applicability of 
Eq. (\ref{maufor}) to quenches within the ferromagnetic phase. 
For all the details about the limit of applicability of Eq. (\ref{maufor}), we refer the reader to Ref. \cite{cef-i}
and here we limit ourselves to the application of this form to cases in which it works.

\subsection{A reduction formula}

For the quench from the ground-state, in Ref. \cite{cef-i} it was very useful to reduce the determination of 
the Pfaffian of $\Gamma$ to the determinant of an $\ell\times \ell$ matrix. 
Although we have not used it in this manuscript, it is worth mentioning that a similar formula holds also for the quench 
from excited states, but the new matrix is more complicated. 
Indeed, the spectral problem for a $2\ell\times 2\ell$ block Toeplitz matrix can be replaced by the spectral problem of an 
$\ell\times \ell$ Toeplitz + Hankel matrix
\begin{equation}
(iT\pm H)\overrightarrow{w_k}=\mp i \lambda_k \overrightarrow{w_k},
\end{equation}
where
\begin{align}
\label{reduced_for_T}
T_{ij}&= \mathrm{sign}(i-j)\sqrt{-h_{i-j}f_{i-j}  },\nonumber\\
H_{ij}&= g_{i+j-\ell-1}.
\end{align}
The matrix $T$ is a Toeplitz matrix and the $H$ has Hankel form. 
Hence we have
\begin{equation}
\rho^{xx}(\ell,t)=\mathrm{pf}(\Gamma)=(-1)^{\frac{\ell(\ell-1)}{2}}\mathrm{det}(H+iT).
\end{equation}
A similar reduction formula can be straightforwardly written down also for the matrix $\Pi$ determining the entanglement entropy.

\end{document}